\documentstyle[12pt,axodraw,graphicx]{article}
\begin{document}
\baselineskip 0.6cm

\def\simgt{\mathrel{\lower2.5pt\vbox{\lineskip=0pt\baselineskip=0pt
           \hbox{$>$}\hbox{$\sim$}}}}
\def\simlt{\mathrel{\lower2.5pt\vbox{\lineskip=0pt\baselineskip=0pt
           \hbox{$<$}\hbox{$\sim$}}}}

\begin{titlepage}

\begin{flushright}
UCB-PTH-08/03
\end{flushright}

\vskip 2.4cm

\begin{center}

{\Large \bf 
Flavorful Supersymmetry from Higher Dimensions
}

\vskip 1.0cm

{\large Yasunori Nomura$^{a,b}$, Michele Papucci$^c$, and 
Daniel Stolarski$^{a,b}$}

\vskip 0.4cm

$^a$ {\it Department of Physics, University of California,
          Berkeley, CA 94720} \\
$^b$ {\it Theoretical Physics Group, Lawrence Berkeley National Laboratory,
          Berkeley, CA 94720} \\
$^c$ {\it School of Natural Sciences, Institute for Advanced Study,
          Princeton, NJ 08540}

\vskip 1.2cm

\abstract{We present models of flavorful supersymmetry in higher 
 dimensions.  The Higgs fields and the supersymmetry breaking field 
 are localized in the same place in the extra dimension(s).  The 
 Yukawa couplings and operators generating the supersymmetry breaking 
 parameters then receive the same suppression factors from the 
 wavefunction profiles of the matter fields, leading to a specific 
 correlation between these two classes of interactions.  The resulting 
 phenomenology is very rich, while stringent experimental constraints 
 from the low-energy flavor and $CP$ violating processes can all be 
 satisfied.  We construct both unified and non-unified models in this 
 framework, which can be either strongly or weakly coupled at the 
 cutoff scale.  We analyze one version in detail, a strongly coupled 
 unified model, which addresses various issues of supersymmetric grand 
 unification.  The models presented here provide an explicit example 
 in which the supersymmetry breaking spectrum can be a direct window 
 into the physics of flavor at a very high energy scale.}

\end{center}
\end{titlepage}

\section{Introduction}
\label{sec:intro}

One of the longstanding puzzles of the standard model is the distinct 
pattern of masses and mixings of the quarks and leptons.  While 
supersymmetry addresses many of the other mysteries of the standard 
model, including the instability of the electroweak scale and the lack 
of a dark matter candidate, it is not clear if and how supersymmetry 
helps us understand the flavor puzzle of the standard model at a deeper 
level.  Recently, it has been pointed out that the supersymmetry 
breaking parameters can exhibit nontrivial flavor structure, and that 
measurement of these parameters at the LHC can give insight into the 
flavor sector of the standard model~\cite{Feng:2007ke,Nomura:2007ap}. 
In particular, it has been shown in Ref.~\cite{Nomura:2007ap} that the 
class of models called flavorful supersymmetry, in which the supersymmetry 
breaking parameters receive similar suppressions to those of the Yukawa 
couplings, can evade all the current experimental bounds and have very 
distinct signatures at the LHC.  In this paper we present explicit 
models of flavorful supersymmetry.

In this paper we construct models in higher dimensional spacetime where 
supersymmetry breaking and the Higgs fields reside in the same location 
in the extra dimension(s).  This provides a simple way to realize the 
necessary correlation between the structures of the supersymmetry breaking 
parameters and the Yukawa couplings~\cite{Kitano:2006ws,Nomura:2007ap}. 
To preserve the successful prediction for supersymmetric gauge coupling 
unification, we take the size of the extra dimension(s) to be of order 
the unification scale.  The hierarchical structure for the Yukawa 
couplings is generated by wavefunction overlaps of the matter and 
Higgs fields~\cite{ArkaniHamed:1999dc}, and the correlation between 
flavor and supersymmetry breaking is obtained by relating the 
location of the Higgs and supersymmetry breaking fields in the 
extra dimension(s).  Models along similar lines were considered 
previously in Ref.~\cite{Kaplan:2000av}, where flavor violation in 
the supersymmetry breaking masses is induced by finite gauge loop 
corrections across the bulk.  Here we consider models in which 
matter fields interact directly with the supersymmetry breaking 
field, giving the simplest scaling for flavorful effects 
in the supersymmetry breaking parameters.%
\footnote{Flavor violation in higher dimensional supersymmetric 
 models was  also discussed in different contexts, 
 see \cite{Hall:2002ci,Abe:2004tq}.}

While not necessary, the extra dimension(s) with size of order the 
unification scale can also be used to address various issues of 
supersymmetric grand unified theories.  Grand unification in higher 
dimensions provides an elegant framework for constructing a simple 
and realistic model of unification~\cite{Kawamura:2000ev,Hall:2001pg}. 
It naturally achieves doublet-triplet splitting in the Higgs sector 
and suppresses dangerous proton decay operators, while preserving 
successful gauge coupling unification.  Realistic quark and lepton 
masses and mixings are also accommodated by placing matter fields 
in the bulk of higher dimensional spacetime~\cite{Hall:2001pg,%
Hall:2001zb,Hebecker:2002re}.  We thus first construct a grand 
unified model of flavorful supersymmetry which can successfully 
address these issues.  In this model we also adopt the assumption 
of strong coupling at the cutoff scale motivated by the 
simplest understanding of gauge coupling unification in higher 
dimensions~\cite{Nomura:2001tn,Hall:2001xb}, although this 
is not a necessity to realize flavorful supersymmetry.

There are a variety of ways to incorporate supersymmetry breaking in 
the present setup.  An important constraint on the flavorful supersymmetry 
framework is that superpotential operators leading to the supersymmetry 
breaking scalar trilinear interactions must be somewhat suppressed, 
unless the superparticles are relatively heavy.  While it is possible 
that this suppression arises accidentally or from physics above the 
cutoff scale, we mainly consider the case where the suppression is 
due to a symmetry under which the supersymmetry breaking field is 
charged.  This symmetry can also be responsible for a complete solution 
to the $\mu$ problem, the problem of the supersymmetric Higgs mass 
term (the $\mu$ term) being of order the weak scale and not some large 
mass scale.  This leads to a scenario similar to the one discussed 
in Refs.~\cite{Ibe:2007km,Nomura:2007cc}, in which the $\mu$ term 
arises from a cutoff suppressed operator~\cite{Giudice:1988yz} 
while the gaugino and sfermion masses are generated by gauge 
mediation~\cite{Dine:1981gu,Dine:1994vc}.  The present setup, 
however, also leads to flavor violating squark and slepton masses 
that are correlated with the Yukawa couplings, characterizing 
flavorful supersymmetry.

We stress that only the extra dimension(s) and the field configuration 
therein are essential for a realization of flavorful supersymmetry. 
All the other ingredients, including grand unification, strong coupling, 
and the particular way of mediating supersymmetry breaking, are not 
important.  While the model described above provides an explicit 
example of flavorful supersymmetry in which many of the issues of 
supersymmetric unification are addressed in a relatively simple setup, 
it is straightforward to eliminate some of the ingredients or to extend 
the model to accommodate more elaborate structures.  In particular, 
we explicitly discuss a construction in which the theory is weakly 
coupled at the cutoff scale, which can be straightforwardly applied 
to models with various spacetime dimensions or gauge groups.

The organization of the paper is as follows.  In the next section we 
present a unified model of flavorful supersymmetry with the assumption 
that the theory is strongly coupled at the cutoff scale.  We explain how 
the relevant correlation between the Yukawa couplings and supersymmetry 
breaking parameters is obtained.  Phenomenology of the model is studied 
in section~\ref{sec:pheno}, including constraints from low-energy 
processes, the superparticle spectrum, and experimental signatures.  In 
section~\ref{sec:warped} we construct a model in warped space, which allows 
us to obtain a picture of realizing flavorful supersymmetry in a 4D setup, 
through the AdS/CFT correspondence.  In section~\ref{sec:weakly-coupled} 
we present a weakly coupled, non-unified model, which does not possess 
a symmetry under which the supersymmetry breaking field is charged. 
Extensions to larger gauge groups or higher dimensions are also discussed. 
Finally, conclusions are given in section~\ref{sec:concl}.

\section{Model}
\label{sec:model}

In this section we present a unified, strongly coupled model.  We 
adopt the simplest setup, $SU(5)$ in 5D, to illustrate the basic idea. 
Extensions to other cases such as larger gauge groups and/or higher 
dimensions are straightforward.  It is also easy to reduce the model 
to a non-unified model in which the gauge group in 5D is the standard 
model $SU(3)_C \times SU(2)_L \times U(1)_Y$.

\subsection{{\boldmath $SU(5)$} grand unification in 5D}
\label{subsec:SU5-5D}

We consider a supersymmetric $SU(5)$ gauge theory in 5D flat spacetime 
with the extra dimension compactified on an $S^1/Z_2$ orbifold: 
$0 \leq y \leq \pi R$, where $y$ represents the coordinate of 
the extra dimension~\cite{Kawamura:2000ev,Hall:2001pg}.  Under 4D 
$N=1$ supersymmetry, the 5D gauge supermultiplet is decomposed into 
a vector superfield $V(A_\mu, \lambda)$ and a chiral superfield 
$\Sigma(\sigma+iA_5, \lambda')$, where both $V$ and $\Sigma$ are 
in the adjoint representation of $SU(5)$.  We impose the following 
boundary conditions on these fields:
\begin{eqnarray}
  && V:\: \left( \begin{array}{ccc|cc}
    (+,+) & (+,+) & (+,+) & (+,-) & (+,-) \\ 
    (+,+) & (+,+) & (+,+) & (+,-) & (+,-) \\ 
    (+,+) & (+,+) & (+,+) & (+,-) & (+,-) \\ \hline
    (+,-) & (+,-) & (+,-) & (+,+) & (+,+) \\ 
    (+,-) & (+,-) & (+,-) & (+,+) & (+,+) 
  \end{array} \right),
\label{eq:bc-V}\\
  && \Sigma:\: \left( \begin{array}{ccc|cc}
    (-,-) & (-,-) & (-,-) & (-,+) & (-,+) \\ 
    (-,-) & (-,-) & (-,-) & (-,+) & (-,+) \\ 
    (-,-) & (-,-) & (-,-) & (-,+) & (-,+) \\ \hline
    (-,+) & (-,+) & (-,+) & (-,-) & (-,-) \\ 
    (-,+) & (-,+) & (-,+) & (-,-) & (-,-) 
  \end{array} \right),
\label{eq:bc-Sigma}
\end{eqnarray}
where $+$ and $-$ represent Neumann and Dirichlet boundary conditions, 
and the first and second signs in parentheses represent boundary 
conditions at $y=0$ and $y=\pi R$, respectively.  This reduces the 
gauge symmetry at $y=\pi R$ to $SU(3) \times SU(2) \times U(1)$, which 
we identify with the standard model gauge group $SU(3)_C \times SU(2)_L 
\times U(1)_Y$ (321).  The zero-mode sector contains only the 321 
component of $V$, $V^{321}$, which is identified with the gauge 
multiplet of the minimal supersymmetric standard model (MSSM).

The Higgs fields are introduced in the bulk as two hypermultiplets 
transforming as the fundamental representation of $SU(5)$.  Using 
notation where a hypermultiplet is represented by two 4D $N=1$ chiral 
superfields $\Phi(\phi,\psi)$ and $\Phi^c(\phi^c,\psi^c)$ with opposite 
gauge transformation properties, our two Higgs hypermultiplets can 
be written as $\{ H, H^c \}$ and $\{ \bar{H}, \bar{H}^c \}$, where 
$H$ and $\bar{H}^c$ transform as ${\bf 5}$ and $\bar{H}$ and $H^c$ 
transform as ${\bf 5}^*$ under $SU(5)$. The boundary conditions are 
given by
\begin{eqnarray}
  H({\bf 5})
    &=& H_T({\bf 3},{\bf 1})_{-1/3}^{(+,-)} 
      \oplus H_D({\bf 1},{\bf 2})_{1/2}^{(+,+)},
\label{eq:bc-H}\\
  H^c({\bf 5}^*)
    &=& H_T^c({\bf 3}^*,{\bf 1})_{1/3}^{(-,+)} 
      \oplus H_D^c({\bf 1},{\bf 2})_{-1/2}^{(-,-)},
\label{eq:bc-Hc}
\end{eqnarray}
for $\{ H, H^c \}$, and similarly for $\{ \bar{H}, \bar{H}^c \}$. 
Here, the right-hand-side shows the decomposition of $H$ and $H^c$ 
into representations of 321 (with $U(1)_Y$ normalized conventionally), 
together with the boundary conditions imposed on each component. 
The zero modes consist of the $SU(2)_L$-doublet components of $H$ 
and $\bar{H}$, $H_D$ and $\bar{H}_D$, which are identified with 
the two Higgs doublets of the MSSM, $H_u$ and $H_d$.

Matter fields are also introduced in the bulk.  To have a complete 
generation, we introduce three hypermultiplets transforming as ${\bf 10}$, 
$\{ T, T^c \}$, $\{ T', T'^c \}$ and $\{ T'', T''^c \}$, two transforming 
as ${\bf 5}^*$, $\{ F, F^c \}$ and $\{ F', F'^c \}$, and one transforming 
as ${\bf 1}$, $\{ O, O^c \}$, for each generation.  The boundary 
conditions are given by
\begin{eqnarray}
  T({\bf 10})
    &=& T_Q({\bf 3},{\bf 2})_{1/6}^{(+,+)} 
      \oplus T_U({\bf 3}^*,{\bf 1})_{-2/3}^{(+,-)} 
      \oplus T_E({\bf 1},{\bf 1})_{1}^{(+,-)},
\label{eq:bc-T}\\
  T'({\bf 10})
    &=& T'_Q({\bf 3},{\bf 2})_{1/6}^{(+,-)} 
      \oplus T'_U({\bf 3}^*,{\bf 1})_{-2/3}^{(+,+)} 
      \oplus T'_E({\bf 1},{\bf 1})_{1}^{(+,-)},
\label{eq:bc-T'}\\
  T''({\bf 10})
    &=& T''_Q({\bf 3},{\bf 2})_{1/6}^{(+,-)} 
      \oplus T''_U({\bf 3}^*,{\bf 1})_{-2/3}^{(+,-)} 
      \oplus T''_E({\bf 1},{\bf 1})_{1}^{(+,+)},
\label{eq:bc-T''}\\
  F({\bf 5}^*)
    &=& F_D({\bf 3}^*,{\bf 1})_{1/3}^{(+,+)} 
      \oplus F_L({\bf 1},{\bf 2})_{-1/2}^{(+,-)},
\label{eq:bc-F}\\
  F'({\bf 5}^*)
    &=& F'_D({\bf 3}^*,{\bf 1})_{1/3}^{(+,-)} 
      \oplus F'_L({\bf 1},{\bf 2})_{-1/2}^{(+,+)},
\label{eq:bc-F'}\\
  O({\bf 1})
    &=& O_N({\bf 1},{\bf 1})_{0}^{(+,+)}.
\label{eq:bc-O}
\end{eqnarray}
The boundary conditions for the conjugated fields are given by 
$+ \leftrightarrow -$, as in Eqs.~(\ref{eq:bc-H},~\ref{eq:bc-Hc}). 
With these boundary conditions, the zero modes arise only from $T_Q$, 
$T'_U$, $T''_E$, $F_D$, $F'_L$ and $O_N$, which we identify with 
a single generation of quark and lepton superfields of the MSSM 
(together with the right-handed neutrino), $Q$, $U$, $E$, $D$, 
$L$ and $N$.%
\footnote{It is possible to extract both $U$ and $E$ from a single 
 hypermultiplet $\{ T', T'^c \}$ by adopting the boundary conditions 
 $T'({\bf 10}) = T'_Q({\bf 3},{\bf 2})_{1/6}^{(+,-)} \oplus T'_U({\bf 
 3}^*,{\bf 1})_{-2/3}^{(+,+)} \oplus T'_E({\bf 1},{\bf 1})_{1}^{(+,+)}$, 
 in which case we do not introduce the hypermultiplet $\{ T'', T''^c \}$. 
 In fact, this is what we obtain if we naively apply the orbifolding 
 procedure to the matter hypermultiplets.  The model also works in 
 this case, with the extra constraint of $M_{U_i} = M_{E_i}$ (see 
 section~\ref{subsec:fermion-masses}) and $q_Q = q_L$ (see 
 section~\ref{subsec:PQ}).}

There are two important scales in the theory: the cutoff scale $M_*$ and 
the compactification scale $1/R$.  We take the ratio of these scales to 
be $\pi R M_* \approx 16\pi^2/g^2 C \approx O(10$~--~$100)$, where $g$ is 
the 4D gauge coupling at the unification scale, $g = O(1)$, and $C \simeq 
5$ is the group theoretical factor for $SU(5)$.  This makes the theory 
strongly coupled at $M_*$, suppressing incalculable threshold corrections 
to gauge coupling unification~\cite{Nomura:2001tn,Hall:2001xb}.%
\footnote{Our estimate on the strong coupling scale is conservative. 
 It is possible that $M_* R$ can be larger by a factor of $\approx \pi$, 
 but it does not affect our results.}
Motivated by successful gauge coupling unification at about 
$10^{16}~{\rm GeV}$ in supersymmetric models, we take the cutoff 
scale and the scale of the extra dimension to be
\begin{equation}
  M_* \approx 10^{17}~{\rm GeV},
\qquad
  1/\pi R \approx 10^{15}~{\rm GeV}.
\label{eq:scales}
\end{equation}
More detailed discussions on gauge coupling unification will be given 
in section~\ref{subsec:gcu}.

\subsection{Quark and lepton masses and mixings}
\label{subsec:fermion-masses}

With the boundary conditions given in the previous subsection, the 
matter content of the theory below $1/R$ reduces to that of the MSSM 
and right-handed neutrinos: $V^{321}$, $H_u$, $H_d$, $Q_i$, $U_i$, $D_i$, 
$L_i$, $E_i$ and $N_i$, where $i=1,2,3$ is the generation index.  The 
Yukawa couplings for the quarks and leptons are introduced on the $y=0$ 
and $y=\pi R$ branes.  The sizes of the 4D Yukawa couplings are then 
determined by the wavefunction values of the matter and Higgs fields 
on these branes.  This can be used to generate the observed hierarchy 
of quark and lepton masses and mixings~\cite{ArkaniHamed:1999dc,%
Kaplan:2000av,Hebecker:2002re}.  Here we consider particular 
configurations of these fields, relevant to our framework.

A nontrivial wavefunction profile for a zero mode can be generated 
by a bulk mass term.  A bulk hypermultiplet $\{ \Phi, \Phi^c \}$ can 
generally have a mass term in the bulk, which is written as
\begin{equation}
  S = \int\!d^4x \int_0^{\pi R}\!\!dy 
      \int\! d^2\theta\, M_\Phi \Phi \Phi^c + {\rm h.c.},
\label{eq:bulk-mass}
\end{equation}
in the basis where the kinetic term is given by $S_{\rm kin} = 
\int\!d^4x \int\!dy\, [\int\!d^4\theta\, (\Phi^\dagger \Phi + \Phi^c 
\Phi^{c\dagger}) + \{ \int\!d^2\theta\, \Phi^c \partial_y \Phi + 
{\rm h.c.} \}]$~\cite{ArkaniHamed:2001tb}.  The wavefunction of 
a zero mode arising from $\Phi$ is proportional to $e^{-M_\Phi y}$, 
so that it is localized to the $y=0$ ($y=\pi R$) brane for 
$M_\Phi > 0$ ($< 0$), and flat for $M_\Phi = 0$.  (The $\Phi^c$ 
case is the same with $M_\Phi \rightarrow -M_\Phi$.)  In the 
present model, we have a bulk mass for each of the Higgs and matter 
hypermultiplets.  For clarity of notation, we specify these masses 
by the subscript representing the corresponding zero mode: $M_{H_u}$, 
$M_{H_d}$, $M_{Q_i}$, $M_{U_i}$, $M_{D_i}$, $M_{L_i}$, $M_{E_i}$ 
and $M_{N_i}$.

We mainly consider the case that the two Higgs doublets $H_u$ and $H_d$ 
are strongly localized to the $y=\pi R$ brane:
\begin{equation}
  M_{H_u},\, M_{H_d} \ll -\frac{1}{R}.
\label{eq:M-Higgs}
\end{equation}
The relevant Yukawa couplings are then those on the $y=\pi R$ brane
\begin{eqnarray}
  S &=& \int\!d^4x \int_0^{\pi R}\!\!dy \,\, 
    \delta(y-\pi R) \int\!d^2\theta\, 
    \biggl\{ (\lambda_u)_{ij} {T_Q}_{i} {T'_U}_j H_D 
\nonumber\\
  && \qquad
      + (\lambda_d)_{ij} {T_Q}_i {F_D}_j \bar{H}_D 
      + (\lambda_e)_{ij} {F'_L}_i {T''_E}_j \bar{H}_D 
      + (\lambda_\nu)_{ij} {F'_L}_i {O_N}_j H_D 
    \biggr\} + {\rm h.c.},
\label{eq:yukawa-321}
\end{eqnarray}
where the sizes of the couplings are naturally given by 
$(\lambda_u)_{ij}, (\lambda_d)_{ij}, (\lambda_e)_{ij}, 
(\lambda_\nu)_{ij} \approx 4\pi/M_*^{3/2}$ using naive 
dimensional analysis~\cite{Manohar:1983md,Nomura:2001tn}. 
This leads to the low-energy 4D Yukawa couplings
\begin{equation}
  W = (y_u)_{ij} Q_i U_j H_u + (y_d)_{ij} Q_i D_j H_d 
    + (y_e)_{ij} L_i E_j H_d + (y_\nu)_{ij} L_i N_j H_u,
\label{eq:4D-Yukawa}
\end{equation}
with
\begin{equation}
  (y_u)_{ij} \approx 4\pi\, \epsilon_{Q_i} \epsilon_{U_j},
\qquad
  (y_d)_{ij} \approx 4\pi\, \epsilon_{Q_i} \epsilon_{D_j},
\qquad
  (y_e)_{ij} \approx 4\pi\, \epsilon_{L_i} \epsilon_{E_j},
\qquad
  (y_\nu)_{ij} \approx 4\pi\, \epsilon_{L_i} \epsilon_{N_j},
\label{eq:Yukawa}
\end{equation}
where the factors $\epsilon_\Phi$ ($\Phi = Q_i, U_i, D_i, L_i, E_i, N_i$) 
are given by
\begin{equation}
  \epsilon_\Phi 
  = \sqrt{\frac{2 M_\Phi}{(1-e^{-2\pi R M_\Phi})M_*}}\, e^{-\pi R M_\Phi} 
  \simeq \left\{ 
  \begin{array}{ll} 
    \sqrt{\frac{2 M_\Phi}{M_*}}\, e^{-\pi R M_\Phi} & 
      {\rm for}\,\,\, \pi R M_\Phi \simgt 1 
\\
    \frac{1}{\sqrt{\pi R M_*}} & 
      {\rm for}\,\,\, |\pi R M_\Phi| \ll 1 
\\
    \sqrt{\frac{2 |M_\Phi|}{M_*}} & 
      {\rm for}\,\,\, \pi R M_\Phi \simlt -1 
  \end{array}. \right.
\label{eq:epsilon}
\end{equation}

Realistic Yukawa couplings are obtained by localizing lighter generations 
more towards the $y=0$ brane so that their wavefunction overlaps with 
the Higgs fields are more suppressed.  For example, we can take
\begin{equation}
\begin{array}{lllll}
  \epsilon_{Q_1} \approx \tilde{y}^{-\frac{1}{2}} \epsilon^2, \quad & 
  \epsilon_{U_1} \approx \tilde{y}^{-\frac{1}{2}} \epsilon^2, \quad & 
  \epsilon_{D_1} \approx \tilde{y}^{-\frac{1}{2}} \epsilon,   \quad & 
  \epsilon_{L_1} \approx \tilde{y}^{-\frac{1}{2}} \epsilon,   \quad & 
  \epsilon_{E_1} \approx \tilde{y}^{-\frac{1}{2}} \epsilon^2,
\\
  \epsilon_{Q_2} \approx \tilde{y}^{-\frac{1}{2}} \epsilon,   \quad & 
  \epsilon_{U_2} \approx \tilde{y}^{-\frac{1}{2}} \epsilon,   \quad & 
  \epsilon_{D_2} \approx \tilde{y}^{-\frac{1}{2}} \epsilon,   \quad & 
  \epsilon_{L_2} \approx \tilde{y}^{-\frac{1}{2}} \epsilon,   \quad & 
  \epsilon_{E_2} \approx \tilde{y}^{-\frac{1}{2}} \epsilon,
\\
  \epsilon_{Q_3} \approx \tilde{y}^{-\frac{1}{2}},            \quad & 
  \epsilon_{U_3} \approx \tilde{y}^{-\frac{1}{2}},            \quad & 
  \epsilon_{D_3} \approx \tilde{y}^{-\frac{1}{2}} \epsilon,   \quad & 
  \epsilon_{L_3} \approx \tilde{y}^{-\frac{1}{2}} \epsilon,   \quad & 
  \epsilon_{E_3} \approx \tilde{y}^{-\frac{1}{2}},
\end{array}
\label{eq:epsilons}
\end{equation}
and
\begin{equation}
  \tan\beta \equiv \frac{\langle H_u \rangle}{\langle H_d \rangle} 
    \approx \epsilon^{-1},
\label{eq:tan-beta}
\end{equation}
where $\epsilon \sim O(0.1)$ and $\tilde{y} \simeq 4\pi \approx 1/\epsilon$, 
to reproduce the gross structure of the observed quark and lepton masses 
and mixings.  The suppression factors of Eq.~(\ref{eq:epsilons}) are 
obtained by taking bulk masses
\begin{equation}
  M_{Q_3, U_3, E_3} \approx -\frac{1}{R}, \qquad 
  M_{Q_2, U_2, D_i, L_i, E_2} \approx \frac{0.5~\mbox{--}~1}{R}, \qquad 
  M_{Q_1, U_1, E_1} \approx \frac{1.5}{R}.
\label{eq:bulk-masses}
\end{equation}
Small neutrino masses are obtained through the seesaw mechanism by 
introducing Majorana masses for the right-handed neutrinos on the 
$y=\pi R$ brane
\begin{equation}
  S = \int\!d^4x \int_0^{\pi R}\!\!dy \,\, 
    \delta(y-\pi R) \int\!d^2\theta\, 
    \frac{(M_N)_{ij}}{2M_*}\, {O_N}_i {O_N}_j + {\rm h.c.}
\label{eq:Majorana}
\end{equation}
The values of $\epsilon_{N_i}$ are then not relevant to the low-energy 
masses and mixings (unless $N_i$'s are localized to the $y=0$ brane 
extremely strongly), since they cancel out in the expression for 
the light neutrino masses.

\begin{figure}
\begin{center}
  \input{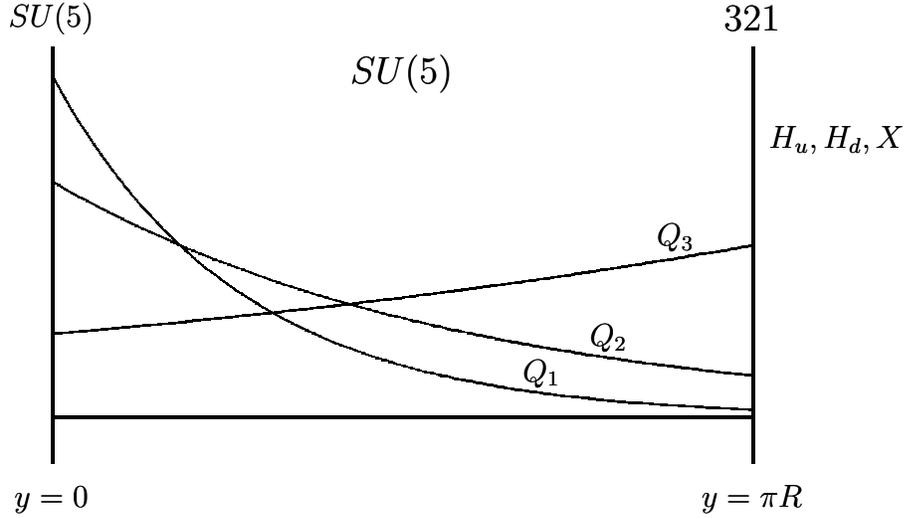}
\caption{A schematic depiction of the localization for various 
 fields.  Here, $X$ represents the supersymmetry breaking field 
 (see section~\ref{subsec:PQ}).}
\label{fig:config}
\end{center}
\end{figure}

The localization of various fields in the extra dimension with the bulk 
masses of Eqs.~(\ref{eq:M-Higgs},~\ref{eq:bulk-masses}) is depicted 
schematically in Fig.~\ref{fig:config}.
The quark and lepton masses and mixings are given by
\begin{equation}
\begin{array}{lll}
  (m_t,m_c,m_u) & \approx & v\, (1,\epsilon^2,\epsilon^4),               \\
  (m_b,m_s,m_d) & \approx & v\, (\epsilon^2,\epsilon^3,\epsilon^4),      \\
  (m_\tau,m_\mu,m_e) & \approx & v\, (\epsilon^2,\epsilon^3,\epsilon^4), \\
  (m_{\nu_\tau},m_{\nu_\mu},m_{\nu_e}) & \approx & \frac{v^2}{M_N} (1,1,1),
\end{array}
\label{eq:q-l-masses}
\end{equation}
and
\begin{equation}
  V_{\rm CKM} \approx 
  \left( \begin{array}{ccc} 
    1 & \epsilon & \epsilon^2 \\
    \epsilon & 1 & \epsilon   \\
    \epsilon^2 & \epsilon & 1
  \end{array} \right),
\qquad
  V_{\rm MNS} \approx 
  \left( \begin{array}{ccc} 
    1 & 1 & 1 \\
    1 & 1 & 1 \\
    1 & 1 & 1
  \end{array} \right),
\label{eq:q-l-mixings}
\end{equation}
where $O(1)$ factors are omitted from each element, and $V_{\rm CKM}$ 
and $V_{\rm MNS}$ are the quark and lepton mixing matrices, respectively. 
This reproduces the gross structure of the observed quark and lepton 
masses and mixings~\cite{Hall:1999sn}.

The matter configuration considered here can be extended easily to 
account for the more detailed pattern of the observed masses and mixings. 
For example, we can localize $L_1$ slightly more towards the $y=0$ brane 
to explain the smallness of the $e3$ element of $V_{\rm MNS}$, which 
is experimentally smaller than about $0.2$.  The other elements of 
$V_{\rm CKM}$ and $V_{\rm MNS}$, as well as the mass eigenvalues, 
can also be better fitted by choosing the bulk masses more carefully. 
Here we simply adopt Eq.~(\ref{eq:bulk-masses}) (and its variations, 
discussed in section~\ref{subsec:flavor}) for the purpose of 
illustrating the general idea.

There are also variations on the location of the Higgs fields.  For 
example, we can localize the two Higgs doublets on the $y=0$ brane, 
instead of the $y = \pi R$ brane: $M_{H_u}, M_{H_d} \simgt 1/\pi R$. 
In this case, the localization should not be very strong so that their 
colored-triplet partners, whose masses are given by $\approx 2 M_{H_u} 
e^{-\pi R M_{H_u}}$ and $2 M_{H_d} e^{-\pi R M_{H_d}}$, do not become 
too light.  The location of the matter fields can simply be flipped 
with respect to $y = \pi R/2$: $M_\Phi \rightarrow -M_\Phi$ for 
$\Phi = Q_i, U_i, D_i, L_i, E_i, N_i$.  Another possibility is to 
(slightly) delocalize $H_u$ and/or $H_d$ from the brane.  In this 
paper, we focus on the case of Eq.~(\ref{eq:M-Higgs}), where $H_u$ 
and $H_d$ are strongly localized to the $y=\pi R$ brane.

\subsection{{\boldmath $\mu$} term, {\boldmath $U(1)_H$}, and flavorful 
 supersymmetry}
\label{subsec:PQ}

In order to have a complete solution to the doublet-triplet splitting 
problem, a possible large mass term for the Higgs doublets on the 
$y=\pi R$ brane, $\delta(y-\pi R) \int\!d^2\theta\, H_D \bar{H}_D$, 
must be forbidden by some symmetry.  Moreover, to understand the weak 
scale size of the mass term ($\mu$ term) for the Higgs doublets, the 
breaking of this symmetry must be associated with supersymmetry breaking. 
One possibility to implement this idea is to consider a $U(1)_R$ 
symmetry under which the two Higgs doublets are neutral~\cite{Hall:2001pg}. 
Here we consider the case that the symmetry is a non-$R$ symmetry.

We consider that the bare $\mu$ term, $\int\!d^2\theta\, H_u H_d$, is 
forbidden, but the effective $\mu$ term is generated by the operator 
$\int\!d^4\theta\, X^\dagger H_u H_d$ through supersymmetry breaking, 
where $X$ is a supersymmetry breaking field~\cite{Giudice:1988yz}.  We 
then find that the relevant symmetry is $U(1)$ (a Peccei-Quinn symmetry) 
whose charge assignment can be taken, without loss of generality, as
\begin{equation}
  Q_i(q_Q),\quad U_i(-1-q_Q),\quad D_i(-1-q_Q),\quad 
  L_i(q_L),\quad E_i(-1-q_L),\quad N_i(-1-q_L),
\label{eq:PQ-matter}
\end{equation}
\begin{equation}
  H_u(1),\quad H_d(1),\quad X(2),
\label{eq:PQ-Higgs}
\end{equation}
where $q_Q$ and $q_L$ are real numbers, and we have assumed that the 
Yukawa couplings are invariant under the symmetry.  In the context 
of the 5D theory, this assignment can be implemented by considering 
$U(1)$ charges for a hypermultiplet $\{ \Phi, \Phi^c \}$ ($\Phi = H, 
\bar{H}, T_i, T'_i, T''_i, F_i, F'_i, O_i$) such that the charge of 
$\Phi$ follows that of the corresponding zero mode, while the charge 
of $\Phi^c$ is the opposite to that of $\Phi$.  This $U(1)$ symmetry 
commutes with 5D supersymmetry.  The $X$ field is introduced on the 
$y=\pi R$ brane, either as a brane field or a bulk field whose zero 
mode is strongly localized to the $y=\pi R$ brane by a bulk mass 
term $M_X \ll -1/R$ (see Fig.~\ref{fig:config}).

The $U(1)$ symmetry of Eqs.~(\ref{eq:PQ-matter},~\ref{eq:PQ-Higgs}), 
which we call $U(1)_H$, has several immediate virtues.  First of all, 
the most general interactions between the Higgs and $X$ fields, located 
on the $y=\pi R$ brane, leads (up to the quadratic order in $X$) to 
the following interactions in 4D:
\begin{equation}
  {\cal L} \approx \int\!d^4\theta\, \left[ 
    \left( \frac{1}{\Lambda} X^\dagger H_u H_d + {\rm h.c.} \right) 
    + \frac{1}{\Lambda^2} X^\dagger X H_u^\dagger H_u 
    + \frac{1}{\Lambda^2} X^\dagger X H_d^\dagger H_d 
    \right],
\label{eq:int-H-X}
\end{equation}
where we have used naive dimensional analysis to estimate the sizes of 
various coefficients, and omitted an $O(1)$ factor in each term.  The 
mass scale $\Lambda$ is defined by
\begin{equation}
  \Lambda \equiv \frac{M_*}{4\pi} \approx 10^{16}~{\rm GeV},
\label{eq:Lambda}
\end{equation}
where we have used Eq.~(\ref{eq:scales}).  After supersymmetry is 
broken by the $F$-term vacuum expectation value (VEV), $F_X$, of the 
$X$ field (see the next subsection), these interactions lead to the 
$\mu$ term and soft supersymmetry breaking masses for the Higgs fields 
of order $F_X/\Lambda$ at the scale $M_*$:
\begin{equation}
  \mu \approx \frac{F_X}{\Lambda},
\qquad
  m_{H_u}^2 \approx m_{H_d}^2 \approx \left(\frac{F_X}{\Lambda}\right)^2.
\label{eq:mu-mH2}
\end{equation}
(Note that $O(1)$ coefficients are omitted in these equations, so that 
the ratio of $m_{H_u}^2$ to $m_{H_d}^2$, for example, can be an arbitrary 
$O(1)$ number.)  An important point here is that the operator ${\cal L} 
\approx \int\!d^4\theta\, (X^\dagger X H_u H_d/\Lambda^2 + {\rm h.c.})$ 
is prohibited by $U(1)_H$, so that the holomorphic supersymmetry breaking 
mass-squared for the Higgs doublets ($B\mu$ term) is not generated at 
order $(F_X/\Lambda)^2$ at tree level.%
\footnote{There are contributions to the $B\mu$ term of order $F_X^2 
 \langle X \rangle/\Lambda^3$ and $F_X m_{3/2}/\Lambda$, where $m_{3/2}$ 
 is the gravitino mass, coming from operators ${\cal L} \approx 
 \int\!d^4\theta\, (X^{\dagger 2} X H_u H_d/\Lambda^3 + {\rm h.c.})$ 
 and the supergravity effects of the first term of Eq.~(\ref{eq:int-H-X}), 
 respectively.  These contributions are, however, negligibly small, 
 since $\langle X \rangle/\Lambda \approx \Lambda/M_{\rm Pl} \ll 1$ and 
 $m_{3/2} \approx F_X/M_{\rm Pl} \ll F_X/\Lambda$, where $M_{\rm Pl} 
 \simeq 2 \times 10^{18}~{\rm GeV}$ is the reduced Planck scale (see 
 section~\ref{subsec:SUSY-br}).}
The low-energy value of the $B\mu$ term is then generated by contributions 
from the gaugino masses through renormalization group evolution.  This 
is crucial to avoid the supersymmetric $CP$ problem, since for weak scale 
superparticle masses an arbitrary relative phase between the $\mu$ and 
$B\mu$ terms leads to an unacceptably large electric dipole moment 
for the electron.

Another important implication of $U(1)_H$ is that possible $y=\pi R$ 
brane operators $\delta(y-\pi R) \int\!d^2\theta\, (X {T_Q}_{i} {T'_U}_j 
H_D + X {T_Q}_i {F_D}_j \bar{H}_D + X {F'_L}_i {T''_E}_j \bar{H}_D 
+ X {F'_L}_i {O_N}_j H_D) + {\rm h.c.}$, which reduce in 4D to 
$\int\!d^2\theta\, (X Q_i U_j H_u + X Q_i D_j H_d + X L_i E_j H_d 
+ X L_i N_j H_u) + {\rm h.c.}$, are forbidden.  If these operators 
were present, they would lead to supersymmetry breaking scalar trilinear 
interactions ($A$ terms) of order $(a_u)_{ij} \approx 4\pi \epsilon_{Q_i} 
\epsilon_{U_j} (F_X/\Lambda)$, $(a_d)_{ij} \approx 4\pi \epsilon_{Q_i} 
\epsilon_{D_j} (F_X/\Lambda)$, $(a_e)_{ij} \approx 4\pi \epsilon_{L_i} 
\epsilon_{E_j} (F_X/\Lambda)$ and $(a_\nu)_{ij} \approx 4\pi \epsilon_{L_i} 
\epsilon_{N_j} (F_X/\Lambda)$, which are not necessarily proportional to 
the corresponding Yukawa matrices in flavor space.  Here, $(a_f)_{ij}$ 
($f=u,d,e,\nu$) are defined by ${\cal L}_{\rm soft} = -(a_u)_{ij} 
\tilde{q}_i \tilde{u}_j h_u - (a_d)_{ij} \tilde{q}_i \tilde{d}_j h_d 
- (a_e)_{ij} \tilde{l}_i \tilde{e}_j h_d - (a_\nu)_{ij} \tilde{l}_i 
\tilde{n}_j h_u + {\rm h.c.}$.  While these terms are suppressed 
by $\epsilon$ factors, they still provide sizable contributions to 
low-energy flavor violating processes, because an $A$-term insertion 
flips the chirality of the sfermion and thus eliminates one factor 
of the Yukawa coupling from an amplitude.  We then find that with 
$O(1)$ coefficients, the rate for $\mu \rightarrow e \gamma$ is expected 
to be larger than the experimental upper bound by a couple of orders 
of magnitude for weak scale superparticle masses~\cite{Nomura:2007ap,%
Kitano:2006ws}.  This problem does not arise in the present model.

The interactions between the matter and $X$ fields relevant to soft 
supersymmetry breaking parameters take the form $\delta(y-\pi R) 
\int\!d^4\theta\, (X^\dagger X {T_Q}_i^\dagger {T_Q}_j + X^\dagger 
X {T'_U}_i^\dagger {T'_U}_j + X^\dagger X {F_D}_i^\dagger {F_D}_j 
+ X^\dagger X {F'_L}_i^\dagger {F'_L}_j + X^\dagger X {T''_E}_i^\dagger 
{T''_E}_j + X^\dagger X {O_N}_i^\dagger {O_N}_j)$, which reduce 
in 4D to
\begin{equation}
  {\cal L} \approx \int\!d^4\theta\, \sum_\Phi \sum_{i,j} 
    \frac{\epsilon_{\Phi_i} \epsilon_{\Phi_j}}{\Lambda^2} 
      X^\dagger X \Phi_i^\dagger \Phi_j,
\label{eq:int-M-X}
\end{equation}
where $\Phi = Q,U,D,L,E,N$.  This leads to the following supersymmetry 
breaking squared masses for the squarks and sleptons at the scale $M_*$:
\begin{equation}
  (m_{\tilde{q}}^2)_{ij} \approx 
    \epsilon_{Q_i} \epsilon_{Q_j} \left(\frac{F_X}{\Lambda}\right)^2,
\qquad
  (m_{\tilde{u}}^2)_{ij} \approx 
    \epsilon_{U_i} \epsilon_{U_j} \left(\frac{F_X}{\Lambda}\right)^2,
\qquad
  (m_{\tilde{d}}^2)_{ij} \approx 
    \epsilon_{D_i} \epsilon_{D_j} \left(\frac{F_X}{\Lambda}\right)^2,
\label{eq:mq2}
\end{equation}
\begin{equation}
  (m_{\tilde{l}}^2)_{ij} \approx 
    \epsilon_{L_i} \epsilon_{L_j} \left(\frac{F_X}{\Lambda}\right)^2,
\qquad
  (m_{\tilde{e}}^2)_{ij} \approx 
    \epsilon_{E_i} \epsilon_{E_j} \left(\frac{F_X}{\Lambda}\right)^2,
\label{eq:ml2}
\end{equation}
where we have omitted supersymmetry breaking masses for the right-handed 
sneutrinos, which are not relevant for low-energy phenomenology.  Through 
Eq.~(\ref{eq:Yukawa}), these masses are related to the Yukawa couplings 
--- lighter generation scalars receive only small contributions, while 
heavier generation scalars can receive sizable ones.  This is exactly 
the pattern needed to realize the flavorful supersymmetry scenario, 
which arises here from the fact that the Higgs and supersymmetry 
breaking fields reside in the same location in the extra dimension. 

As shown in Ref.~\cite{Nomura:2007ap}, the existence of flavor non-universal 
contributions of Eqs.~(\ref{eq:mq2},~\ref{eq:ml2}) does not contradict 
the low-energy data on flavor or $CP$ violating processes for wide 
parameter regions.  Since the masses of Eqs.~(\ref{eq:mq2},~\ref{eq:ml2}) 
are highly flavor non-universal, they cannot be the dominant contribution 
to the soft masses (except possibly for some of the third generation 
sfermions), and we need an extra flavor universal contribution as 
well as the gaugino masses.  These are generated in the present model 
by gauge mediation, as discussed in the next subsection.

Finally, the $U(1)_H$ symmetry forbids any superpotential term involving 
only the $X$ field.  Since breaking supersymmetry requires a linear $X$ 
term in the superpotential, this implies that supersymmetry is not broken 
unless $U(1)_H$ is broken, providing a solid relation between breaking 
of supersymmetry and that of $U(1)_H$.

\subsection{Supersymmetry breaking and the low-energy spectrum}
\label{subsec:SUSY-br}

To induce supersymmetry breaking VEV $F_X$, we need a linear term of 
$X$ in the superpotential.  This implies that $U(1)_H$ must be broken 
either explicitly or spontaneously.  Here we simply parameterize the 
effect of $U(1)_H$ breaking in the $X$ potential by a dimensionless 
chiral spurious parameter $\eta$, which we assume to have the $U(1)_H$ 
charge of $-2$.  The resulting physics does not depend much on the 
underlying origin of this breaking.

The most general low-energy 4D interactions of $X$ consistent with the 
(broken) $U(1)_H$ symmetry is given by the following K\"ahler potential 
and superpotential:
\begin{equation}
  K \approx X^\dagger X 
    - \frac{1}{4\Lambda^2} (X^\dagger X)^2 + \cdots,
\label{eq:K_X}
\end{equation}
\begin{equation}
  W \approx c + \mu_X^2 X + \frac{\mu_X^4}{4\pi \Lambda^3} X^2 
      + \frac{\mu_X^6}{(4\pi)^2 \Lambda^6} X^3 + \cdots,
\label{eq:W_X}
\end{equation}
where $c$ is a constant term in the superpotential, needed to cancel the 
cosmological constant, and $\mu_X^2 \equiv 4\pi \eta \Lambda^2$.  Here, 
again, we have used naive dimensional analysis to estimate the sizes 
of various coefficients (except for the $c$ term), and omitted an $O(1)$ 
factor in each term.%
\footnote{The most general insertions of the spurious parameter $\eta$ 
 allows us to write down the tree-level $\mu$ term in the superpotential, 
 with $\mu \approx 4\pi \eta \Lambda \approx \mu_X^2/\Lambda$.  This 
 contribution is the same order as the one in Eq.~(\ref{eq:mu-mH2}); 
 see Eq.~(\ref{eq:X-min}).}
Note that the terms in Eqs.~(\ref{eq:K_X},~\ref{eq:W_X}) arise from 
operators localized on the $y=\pi R$ brane, except for the $c$ term 
which can have contributions from other sources as well.

The scalar potential arising from Eqs.~(\ref{eq:K_X},~\ref{eq:W_X}) 
can be minimized in supergravity.  Assuming that the coefficient of 
the $(X^\dagger X)^2/\Lambda^2$ term in the K\"ahler potential is 
negative, the minimum of $X$ is given by the competition between the 
$X$ mass term arising from $V \simeq (\mu_X^4/\Lambda^2) |X|^2 \subset 
|\partial W/\partial X|^2 (\partial^2 K/\partial X^\dagger \partial X)^{-1}$ 
and the linear term $V \simeq -2\mu_X^2 c (X+X^\dagger)/M_{\rm Pl}^2$ 
arising in supergravity.  The constant $c$ is determined to cancel the 
vacuum energy $V \simeq |\partial W/\partial X|^2 - 3|W|^2/M_{\rm Pl}^2$ 
as $c \simeq \mu_X^2 M_{\rm Pl}/\sqrt{3}$.  This, therefore, leads to 
the following supersymmetry breaking minimum
\begin{equation}
  \langle X \rangle \simeq \frac{2 \Lambda^2}{\sqrt{3} M_{\rm Pl}} 
    \approx 10^{14}~{\rm GeV},
\qquad
  F_X \simeq \mu_X^2,
\label{eq:X-min}
\end{equation}
with the mass-squared for the $X$ excitation given by $m_X^2 \approx 
\mu_X^4/\Lambda^2$.  Note that the $X$ VEV, $\langle X \rangle \approx 
10^{14}~{\rm GeV}$, is smaller than the compactification scale, $1/\pi R 
\approx 10^{15}~{\rm GeV}$, so that the 4D analysis of the potential 
minimization is justified.  In fact, with $\mu_X$ much smaller than 
$\langle X \rangle$ to reproduce the weak scale superparticle masses 
(see Eqs.~(\ref{eq:mu-mH2},~\ref{eq:mq2},~\ref{eq:ml2}) and below), 
the only relevant terms in the potential minimization are the first 
two terms of Eqs.~(\ref{eq:K_X}) and (\ref{eq:W_X}).

The supersymmetry breaking of Eq.~(\ref{eq:X-min}) can be transmitted 
to the MSSM gauginos and scalars by gauge mediation by coupling 
$X$ to the messenger fields $f$ and $\bar{f}$: $W = \lambda X f 
\bar{f}$~\cite{Kitano:2006wz}.  The minimum of $X$ in Eq.~(\ref{eq:X-min}) 
is not destabilized as long as the coupling $\lambda$ is sufficiently 
small, $\lambda^2 n_f/16\pi^2 \simlt (\Lambda/M_{\rm Pl})^2$, where $n_f$ 
is the number of components for the messenger fields.  We introduce the 
messenger fields in the bulk as hypermultiplets: $\{ f, f^c \}$ and 
$\{ \bar{f}, \bar{f}^c \}$.  The boundary conditions are given by
\begin{eqnarray}
  f({\bf 5})
    &=& f_D({\bf 3},{\bf 1})_{-1/3}^{(+,+)} 
      \oplus f_L({\bf 1},{\bf 2})_{1/2}^{(+,+)},
\label{eq:bc-f}\\
  f^c({\bf 5}^*)
    &=& f_D^c({\bf 3}^*,{\bf 1})_{1/3}^{(-,-)} 
      \oplus f_L^c({\bf 1},{\bf 2})_{-1/2}^{(-,-)},
\label{eq:bc-fc}
\end{eqnarray}
and similarly for $\{ \bar{f}, \bar{f}^c \}$, leading to the zero modes 
from $f_D$, $f_L$, $\bar{f}_D$ and $\bar{f}_L$.  Here, we have chosen 
the messenger fields to be a pair of ${\bf 5} + {\bf 5}^*$, for simplicity, 
but they can in general be an arbitrary number of pairs of arbitrary 
$SU(5)$ representations (as long as they do not make the standard model 
gauge couplings strong at or below $\sim 1/R$).  The messenger fields 
have interactions to $X$ on the $y=\pi R$ brane:
\begin{equation}
  S = \int\!d^4x \int_0^{\pi R}\!\!dy \,\, 
    \delta(y-\pi R) \int\!d^2\theta\, (\eta_D\, X f_D \bar{f}_D 
      + \eta_L\, X f_L \bar{f}_L) + {\rm h.c.},
\label{eq:Xffbar}
\end{equation}
where the couplings $\eta_D$ and $\eta_L$ are of order $4\pi/M_*$ 
($4\pi/M_*^{3/2}$) from naive dimensional analysis if $X$ is a $y=\pi R$ 
brane (bulk) field.  This determines the $U(1)_H$ charges of the 
$f = f_D + f_L$ and $\bar{f} = \bar{f}_D + \bar{f}_L$ fields such 
that the sum of the $f$ and $\bar{f}$ charges is $-2$. (The $f^c$ 
and $\bar{f}^c$ fields have the opposite charges to $f$ and $\bar{f}$, 
respectively.)

The messenger multiplets in general have the bulk mass terms of the form 
of Eq.~(\ref{eq:bulk-mass}), $M_f$ and $M_{\bar{f}}$.  The interactions 
of Eq.~(\ref{eq:Xffbar}) then lead to the 4D superpotential
\begin{equation}
  W = \lambda_D X f_D \bar{f}_D + \lambda_L X f_L \bar{f}_L,
\label{eq:W_Xffbar}
\end{equation}
where $f_D$, $f_L$, $\bar{f}_D$ and $\bar{f}_L$ represent the zero-mode 
chiral superfields, and
\begin{equation}
  \lambda_D \approx \lambda_L \approx 4\pi\, \epsilon_f \epsilon_{\bar{f}},
\label{eq:lambda}
\end{equation}
where $\epsilon_f, \epsilon_{\bar{f}}$ are given by Eq.~(\ref{eq:epsilon}) 
with $\Phi = f, \bar{f}$.  The stability condition for the potential is 
$\lambda_{D,L}^2 n_f/16\pi^2 \simlt (\Lambda/M_{\rm Pl})^2 \approx 10^{-4}$, 
which can be easily satisfied, for example, by taking $M_f, M_{\bar{f}} 
\simgt 1/\pi R$, i.e., $f_D$, $f_L$, $\bar{f}_D$ and $\bar{f}_L$ 
localized towards the $y=0$ brane.  At the scale
\begin{equation}
  M_{\rm mess} \approx \lambda_{D,L} \langle X \rangle 
    \approx \frac{\lambda_{D,L} \Lambda^2}{M_{\rm Pl}},
\label{eq:M_mess}
\end{equation}
the messenger fields are integrated out, generating the 
gauge-mediated contributions to the MSSM gaugino and scalar 
masses~\cite{Dine:1981gu,Dine:1994vc}:
\begin{equation}
  M_a = N_{\rm mess} \frac{g_a^2}{16\pi^2} \frac{F_X}{\langle X \rangle},
\qquad
  m_{\tilde{f}}^2 = 2 N_{\rm mess} \sum_a C_a^{\tilde{f}} 
      \left( \frac{g_a^2}{16\pi^2} \right)^2 
      \left| \frac{F_X}{\langle X \rangle} \right|^2,
\label{eq:GMSB}
\end{equation}
where $a = 1,2,3$ represents the standard model gauge group factors, 
$g_a$ are the standard model gauge couplings at $M_{\rm mess}$, 
$\tilde{f} = \tilde{q}, \tilde{u}, \tilde{d}, \tilde{l}, \tilde{e}, 
H_u, H_d$, and $C_a^{\tilde{f}}$ are the quadratic Casimir 
coefficients.

The supersymmetry breaking parameters and the $\mu$ parameter 
in our theory receive contributions of Eqs.~(\ref{eq:mu-mH2},%
~\ref{eq:mq2},~\ref{eq:ml2}) generated at the scale $M_*$ and those 
of Eq.~(\ref{eq:GMSB}) generated at the scale $M_{\rm mess}$.  The 
low-energy superparticle masses are then obtained by evolving the 
parameters of Eqs.~(\ref{eq:mu-mH2},~\ref{eq:mq2},~\ref{eq:ml2}) 
from $M_*$ to $M_{\rm mess}$, adding the contributions of 
Eq.~(\ref{eq:GMSB}) at $M_{\rm mess}$, and then evolving the 
resulting parameters from $M_{\rm mess}$ down to the weak scale. 
Because of the wavefunction suppression factors $\epsilon_{f,\bar{f}}$, 
which are exponentially sensitive to the bulk masses $M_{f,\bar{f}}$, 
the value of $M_{\rm mess}$ can in general be anywhere between 
$\approx 100~{\rm TeV}$ and $O(0.1)\langle X \rangle \approx 
10^{13}~{\rm GeV}$.  Here, the upper bound comes from the stability 
condition on $\lambda_{D,L}$, while the lower bound from the messenger 
stability.  Note that since the gauge-mediated contributions of 
Eq.~(\ref{eq:GMSB}) have the size
\begin{equation}
  M_a \approx (m_{\tilde{f}}^2)^{1/2} 
  \approx \frac{F_X}{\Lambda} 
    \left( \frac{g^2}{16\pi^2} \frac{M_{\rm Pl}}{\Lambda} \right) 
  \approx \frac{F_X}{\Lambda},
\label{eq:GMSB-size}
\end{equation}
where $g$ represents the standard model gauge couplings, they are 
comparable to the tree-level contributions to the Higgs-sector parameters 
of Eq.~(\ref{eq:mu-mH2}).%
\footnote{In contrast with the situation discussed in Ref.~\cite{Ibe:2007km}, 
 there is no reason in the present theory that the $\mu$ term must be 
 suppressed compared with the gauge-mediated contributions.  In fact, 
 they are naturally expected to be comparable.}
On the other hand, the flavor non-universal contributions of 
Eqs.~(\ref{eq:mq2},~\ref{eq:ml2}) are suppressed due to the $\epsilon$ 
factors associated with the quark and lepton superfields (except possibly 
for the third generation).  This therefore reproduces precisely the 
pattern for the low-energy supersymmetry breaking masses in flavorful 
supersymmetry.

The model also has other flavor violating contributions to the 
supersymmetry breaking parameters, but they are all small.  For example, 
loops of the higher dimensional gauge and messenger fields produce 
flavor violating scalar squared masses at $1/R$, but they are of 
order $N_{\rm mess}(g^2/16\pi^2)^2 |F_X/\Lambda|^2 \approx (\langle 
X \rangle/\Lambda)^2 m_{\tilde{f}}^2$ and thus small.  The $y=0$ brane 
K\"ahler potential operators connecting the matter (and messenger) 
fields, e.g. $\delta(y) \int\!d^4\theta\, T_i^\dagger T_j T_k^\dagger 
T_l$ and $\delta(y) \int\!d^4\theta\, T_i^\dagger T_j f^\dagger f$, 
also generate flavor violating scalar squared masses through loops of 
the matter (or messenger) fields.  Using naive dimensional analysis 
to estimate the coefficients of the operators, we find that this 
contribution is at most of order $|F_X/\Lambda|^2/(\pi R M_*)^5$ 
and negligible.  Possible contributions from bulk higher dimension 
operators are also expected to be small based on similar dimensional 
arguments.  Finally, $y=0$ brane localized kinetic terms, e.g. 
$\delta(y) \int\!d^4\theta\, T_i^\dagger T_j$, can introduce flavor 
violation by giving corrections of order $1/M_* R \approx 1/16\pi^2$ 
to the kinetic terms of the low energy 4D fields.  After canonically 
normalizing the 4D fields, these corrections affect both the Yukawa 
couplings and the supersymmetry breaking parameters.  Interestingly, 
however, this does not affect the mass insertion parameters used in 
section~\ref{subsec:flavor} at the order of magnitude level.  In 
other words, we can always take the basis for the low energy 4D fields 
such that the Yukawa couplings and supersymmetry breaking masses are 
given by Eqs.~(\ref{eq:Yukawa},~\ref{eq:mq2},~\ref{eq:ml2}) at $M_*$ 
even in the presence of the general brane kinetic terms.%
\footnote{In fact, this property persists even if the corrections to 
 the 4D kinetic terms are of order unity.}
Below, we assume that this basis is taken.

Setting the size of the dominant contributions to the supersymmetry 
breaking and $\mu$ parameters to be the weak scale, we obtain $F_X/\Lambda 
\approx (100~{\rm GeV}$~--~$1~{\rm TeV})$ from Eq.~(\ref{eq:GMSB-size}). 
The value of $F_X$ is then determined as $\sqrt{F_X} \approx 
(10^{8.5}$~--~$10^{9.5})~{\rm GeV}$ using Eq.~(\ref{eq:Lambda}). 
This leads to the gravitino mass
\begin{equation}
  m_{3/2} \simeq \frac{F_X}{\sqrt{3}M_{\rm Pl}} 
    \approx (0.1~\mbox{--}~10)~{\rm GeV},
\label{eq:m32}
\end{equation}
implying that the gravitino is the lightest supersymmetric particle 
(LSP).  Together with the flavor non-universal contributions of 
Eqs.~(\ref{eq:mq2},~\ref{eq:ml2}), this can lead to spectacular 
signatures at the LHC~\cite{Nomura:2007ap}, some of which will 
be discussed in section~\ref{subsec:signature}.

\subsection{Neutrino masses, {\boldmath $R$} parity, and 
 dimension five proton decay}
\label{subsec:U1H}

The $U(1)_H$ charge assignment of Eqs.~(\ref{eq:PQ-matter},%
~\ref{eq:PQ-Higgs}) contains two free parameters $q_Q$ and $q_L$. 
These parameters can be restricted by imposing various phenomenological 
requirements~\cite{Nomura:2007cc}.  For example, if we require that 
dangerous dimension-five proton decay operators $W \sim Q_i Q_j Q_k L_l$ 
and $U_i U_j D_k E_l$ are prohibited by $U(1)_H$, then we obtain the 
conditions $3q_Q+q_L \neq 0$ and $3q_Q+q_L \neq -4$, respectively. 
Similarly, if we require that $U(1)_H$ forbids dimension-four $R$-parity 
violating operators $W \sim L_i H_u$, $Q_i D_j L_k$, $U_i D_j D_k$, 
$L_i L_j E_k$ and $K \sim L_i^\dagger H_d$, we obtain $q_L \neq -1$, 
$q_L \neq 1$, $q_Q \neq -1$, $q_L \neq 1$ and $q_L \neq 1$.

An interesting possibility arises if $q_L=0$.  In this case we can have 
the following superpotential on the $y=\pi R$ brane:
\begin{equation}
  S = \int\!d^4x \int_0^{\pi R}\!\!dy \,\, 
    \delta(y-\pi R) \int\!d^2\theta\, 
      \frac{\hat{\kappa}_{ij}}{2} X {O_N}_{i} {O_N}_j + {\rm h.c.},
\label{eq:seesaw}
\end{equation}
which, together with the last term of Eq.~(\ref{eq:yukawa-321}), leads to
\begin{equation}
  W = \frac{\kappa_{ij}}{2} X N_i N_j 
    + (y_\nu)_{ij} L_i N_j H_u,
\label{eq:4D-seesaw}
\end{equation}
in the low-energy 4D theory.  Using naive dimensional analysis, the 
couplings $\kappa_{ij}$ and $(y_\nu)_{ij}$ are given by $\kappa_{ij} 
\approx 4\pi\, \epsilon_{N_i} \epsilon_{N_j}$ and $(y_\nu)_{ij} 
\approx 4\pi\, \epsilon_{L_i} \epsilon_{N_j}$.  The vacuum of 
Eq.~(\ref{eq:X-min}) is not destabilized as long as $\kappa_{ij} 
\simlt O(0.1)$, which can be easily satisfied by taking $\epsilon_{N_i}$
 to be somewhat small, i.e., by taking $M_{N_i} \simgt -1/\pi R$. 
Small neutrino masses are then generated by the seesaw mechanism 
through the $X$ VEV of Eq.~(\ref{eq:X-min}).  Note that the 
$\epsilon_{N_i}$ factors cancel out from the generated neutrino 
masses:
\begin{equation}
  (m_\nu)_{ij} \approx 4\pi \epsilon_{L_i} \epsilon_{L_j}
    \frac{\langle H_u \rangle^2}{\langle X \rangle}.
\label{eq:nu-mass}
\end{equation}
It is interesting that with $\langle X \rangle \approx 10^{14}~{\rm GeV}$, 
this is in the right ballpark to explain the experimental data on 
neutrino oscillations.%
\footnote{The interactions of Eq.~(\ref{eq:4D-seesaw}) also generate 
 supersymmetry breaking masses of order $(y_\nu^2/16\pi^2)F_X/\langle 
 X \rangle$ for $L_i$ and $H_u$ through loops of $N_i$ ($A$ terms 
 at one loop and non-holomorphic supersymmetry breaking masses 
 at two loops~\cite{Giudice:1997ni}).  This effect, however, is 
 small for $y_\nu \ll 1$, compared with the contributions of 
 Eqs.~(\ref{eq:mu-mH2},~\ref{eq:GMSB}).}

It is not necessary to impose all the requirements above for the 
$U(1)_H$ charge assignment.  For example, $R$-parity violating operators 
can be forbidden simply by imposing matter parity in addition to $U(1)_H$. 
Nevertheless, it is interesting that one can consider the $U(1)_H$ 
assignment that satisfies all these requirements.  For example, 
one can adopt
\begin{equation}
  q_Q = \frac{4}{3} + 2n,
\qquad
  q_L = 0,
\label{eq:x-y}
\end{equation}
where $n$ is an integer.  The $U(1)_H$ symmetry is spontaneously broken 
by the VEV of $X$, but the charge assignment of Eq.~(\ref{eq:x-y}) 
leaves a discrete $Z_6$ symmetry after the breaking.  The product of 
$Z_6$ and $U(1)_Y$ contains the (anomalous) $Z_3$ baryon number and 
(anomaly-free) $Z_2$ matter parity ($R$ parity) as subgroups.  This 
symmetry, therefore, strictly forbids the $R$-parity violating operators, 
and the lightest supersymmetric particle is absolutely stable.

In the rest of the paper, we assume that the LSP is absolutely stable 
(although it is not necessarily required by the model).  This can 
be achieved either by choosing the $U(1)_H$ charges so that all the 
$R$-parity violating operators are forbidden even after the $U(1)_H$ 
breaking, as is the case for Eq.~(\ref{eq:x-y}), or simply by imposing 
matter (or $R$) parity.

\subsection{Origin of {\boldmath $U(1)_H$} breaking}
\label{subsec:U1H-breaking}

In section~\ref{subsec:SUSY-br}, we have simply parameterized the 
effect of (small) $U(1)_H$ breaking by a spurious parameter $\eta \ll 1$. 
This breaking controls the size of the coefficient $\mu_X^2$ for the 
$X$ linear term in the superpotential, and thus the size of supersymmetry 
breaking.  There are a variety of possibilities for the origin of 
the required small breaking.  For example, it may simply arise as 
a result of string theory dynamics at the cutoff scale $M_*$.  Here, 
we discuss two explicit examples for the origin of $U(1)_H$ breaking. 
The validity of the model as well as its basic phenomenological 
consequences discussed in section~\ref{sec:pheno} have little 
dependence on this physics.

The first possibility is that the $U(1)_H$ breaking effect arises 
from the mixed $U(1)_H$ anomaly with respect to the hidden sector gauge 
group.  The scale $\mu_X$ then arises from dimensional transmutation 
associated with the hidden sector gauge group.  This scenario can be 
implemented in our higher dimensional framework simply by promoting 
the model discussed in Refs.~\cite{Nomura:2007cc,Ibe:2007gf} to 
higher dimensions.  Specifically, we consider a supersymmetric 
$SU(5)_{\rm hid} \times SU(5)$ gauge theory on 5D flat spacetime, 
where the latter $SU(5)$ factor is identified with the unified 
gauge group, whose gauge multiplet obeys the boundary conditions of 
Eqs.~(\ref{eq:bc-V},~\ref{eq:bc-Sigma}).  The Higgs and matter fields 
are singlet under $SU(5)_{\rm hid}$, and have the same $SU(5)$ gauge 
quantum numbers and boundary conditions as in section~\ref{subsec:SU5-5D}. 
The location for the Higgs, matter and $X$ fields, as well as their 
$U(1)_H$ charges, are also the same as before.

The messenger fields $\{ f, f^c \}$ and $\{ \bar{f}, \bar{f}^c \}$ 
are also introduced in the bulk as before, with the interactions to 
the $X$ field given by Eq.~(\ref{eq:Xffbar}).  Instead of introducing 
arbitrary explicit $U(1)_H$ breaking, however, here we assign the gauge 
quantum numbers $({\bf 5}^*, {\bf 5})$ to $f$ and $\bar{f}^c$, and 
$({\bf 5}, {\bf 5}^*)$ to $\bar{f}$ and $f^c$, where the numbers in 
parentheses represent the quantum numbers under $SU(5)_{\rm hid} \times 
SU(5)$.  Below the compactification scale $\approx 1/\pi R$, this 
reduces to the model discussed in~\cite{Nomura:2007cc,Ibe:2007gf}. 
In particular, the required $X$ linear term in the superpotential 
is generated:
\begin{equation}
  W_{\rm eff} = \lambda \Lambda_{\rm hid}^2 X,
\label{eq:dyn-X-lin}
\end{equation}
where $\Lambda_{\rm hid}$ is the dynamical scale of $SU(5)_{\rm hid}$, 
and we have taken $\lambda_D \approx \lambda_L \approx \lambda$. 
Note that this superpotential is ``exact,'' i.e., no higher order 
terms in $X$ are generated.

A virtue of the higher dimensional setup in the context of 
$SU(5)_{\rm hid} \times SU(5)$ is that the nontrivial wavefunction 
profiles of $f$ and $\bar{f}$ needed to suppress $\lambda_{D,L}$ 
(to satisfy the stability condition $\lambda_{D,L}^2 \simlt 10^{-3}$) 
also suppress the superpotential coupling $W = \zeta f\bar{f}H_u 
H_d/\Lambda$ in the low-energy 4D theory, which can arise from the 
$y=\pi R$ brane localized operator and leads to an unwanted large 
$\mu$ term unless $\zeta \simlt \lambda_{D,L}$.  Using naive dimensional 
analysis, we find $\lambda_{D,L} \approx \zeta \approx 4\pi \epsilon_f 
\epsilon_{\bar{f}}$, so that we do not have a large $\mu$ term 
from the superpotential operator.

Another possibility for the $U(1)_H$ breaking is that $U(1)_H$ 
is spontaneously broken.  Since $U(1)_H$ has a mixed anomaly with 
respect to $SU(3)_C$, this provides a solution to the strong $CP$ 
problem~\cite{Peccei:1977hh}.  We do not attempt here to construct 
a complete model of this kind.  It is, however, straightforward to 
realize this possibility at the level of a non-linear sigma model, 
i.e. the axion field being realized nonlinearly.

\section{Phenomenology}
\label{sec:pheno}

In this section we study phenomenology of the model presented in the 
previous section.  We study constraints from flavor and $CP$ violation 
and the variation of the superparticle spectrum allowed by these 
constraints.  We find that there are a variety of possibilities for 
the next-to-lightest supersymmetric particle (NLSP), which decays into 
the LSP gravitino with the lifetime of $O(10^2$~--~$10^6~{\rm sec})$. 
We also discuss proton decay, precision gauge coupling unification, 
and possible experimental signatures.

\subsection{Constraints from flavor violation and the variety of the spectrum}
\label{subsec:flavor}

Phenomenology of the model depends on the wavefunction profiles for 
the quark and lepton zero modes, which are controlled by the bulk masses 
for these fields.  In the low-energy 4D theory, these affect the Yukawa 
matrices, Eq.~(\ref{eq:Yukawa}), and the flavor violating contribution to 
the squark and slepton masses generated at $M_*$, Eqs.~(\ref{eq:mq2},%
~\ref{eq:ml2}).  This effect is parameterized by the factors $\epsilon_\Phi$ 
($\Phi = Q_i, U_i, D_i, L_i, E_i, N_i$) in Eq.~(\ref{eq:epsilon}).

The values for the $\epsilon_\Phi$ factors are restricted by requiring 
that the gross structure of the observed quark and lepton masses and 
mixings are reproduced by these factors.  This, however, still leaves 
some freedoms for the choice of the $\epsilon_\Phi$ factors.  For example, 
scaling $\{ \epsilon_{Q_i}, \epsilon_{U_i}, \epsilon_{D_i} \} \rightarrow 
\{ \alpha \epsilon_{Q_i}, \alpha^{-1} \epsilon_{U_i}, \alpha^{-1} 
\epsilon_{D_i} \}$ does not change the quark masses and mixings. 
Taking these freedoms into account, here we consider
\begin{equation}
\begin{array}{lll}
  \epsilon_{Q_1} \approx 
    \tilde{y}^{-\frac{1}{2}} \alpha_q\, \epsilon^2,                 \quad & 
  \epsilon_{U_1} \approx 
    \tilde{y}^{-\frac{1}{2}} \alpha_q^{-1} \epsilon^2,              \quad & 
  \epsilon_{D_1} \approx 
    \tilde{y}^{-\frac{1}{2}} \alpha_q^{-1} \alpha_\beta\, \epsilon,
\\
  \epsilon_{Q_2} \approx 
    \tilde{y}^{-\frac{1}{2}} \alpha_q\, \epsilon,                   \quad & 
  \epsilon_{U_2} \approx 
    \tilde{y}^{-\frac{1}{2}} \alpha_q^{-1} \epsilon,                \quad & 
  \epsilon_{D_2} \approx 
    \tilde{y}^{-\frac{1}{2}} \alpha_q^{-1} \alpha_\beta\, \epsilon,
\\
  \epsilon_{Q_3} \approx 
    \tilde{y}^{-\frac{1}{2}}  \alpha_q,                             \quad & 
  \epsilon_{U_3} \approx 
    \tilde{y}^{-\frac{1}{2}} \alpha_q^{-1},                         \quad & 
  \epsilon_{D_3} \approx 
    \tilde{y}^{-\frac{1}{2}} \alpha_q^{-1} \alpha_\beta\, \epsilon,
\end{array}
\label{eq:epsilon-QUD}
\end{equation}
\begin{equation}
\begin{array}{ll}
  \epsilon_{L_1} \approx 
    \tilde{y}^{-\frac{1}{2}} \alpha_l\, \epsilon,                     \quad & 
  \epsilon_{E_1} \approx 
    \tilde{y}^{-\frac{1}{2}} \alpha_l^{-1} \alpha_\beta\, \epsilon^2,
\\
  \epsilon_{L_2} \approx 
    \tilde{y}^{-\frac{1}{2}} \alpha_l\, \epsilon,                     \quad & 
  \epsilon_{E_2} \approx 
    \tilde{y}^{-\frac{1}{2}} \alpha_l^{-1} \alpha_\beta\, \epsilon,
\\
  \epsilon_{L_3} \approx 
    \tilde{y}^{-\frac{1}{2}}  \alpha_l\, \epsilon,                    \quad & 
  \epsilon_{E_3} \approx 
    \tilde{y}^{-\frac{1}{2}} \alpha_l^{-1} \alpha_\beta,
\end{array}
\label{eq:epsilon-LE}
\end{equation}
with
\begin{equation}
  \tan\beta \approx \alpha_\beta\, \epsilon^{-1},
\label{eq:tan-beta-gen}
\end{equation}
where $\epsilon = O(0.1)$ and $\alpha_q$, $\alpha_l$ and $\alpha_\beta$ 
are numbers parameterizing the freedoms unfixed by the data of 
the quark and lepton masses and mixings.  Note that the range of 
$\alpha_{q,l,\beta}$ is restricted such that the $\epsilon$ parameters, 
$\epsilon_{Q_i,U_i,D_i,L_i,E_i}$, do not exceed $\approx 1$; see 
Eq.~(\ref{eq:epsilon}).  (The value of $\alpha_\beta$ is also restricted 
so that $\tan\beta$ stays within the regime in which none of the 
Yukawa couplings blow up below the cutoff scale.)  The pattern of 
Eqs.~(\ref{eq:epsilon-QUD},~\ref{eq:epsilon-LE}) is a straightforward 
generalization of Eq.~(\ref{eq:epsilons}), and the resulting 
quark and lepton masses and mixings are still given by 
Eqs.~(\ref{eq:q-l-masses},~\ref{eq:q-l-mixings}).

The parameters  $\alpha_{q}$, $\alpha_{l}$ and $\alpha_{\beta}$, 
however, alter the size of the flavor violating contribution to the 
squark and slepton masses, and are thus constrained by low-energy 
flavor and $CP$ violating processes.  We use the mass insertion 
method~\cite{Hall:1985dx} to derive constraints on these parameters. 
The experimental bounds on the mass insertion parameters can be 
found, e.g., in Ref.~\cite{Gabbiani:1996hi}, and are summarized 
in Ref.~\cite{Nomura:2007ap}.  In the quark sector, the most stringent 
bounds come from $K$-$\bar{K}$, $D$-$\bar{D}$ and $B$-$\bar{B}$ mixings 
and $\sin 2\beta$, while in the lepton sector the most stringent one 
comes from the $\mu \rightarrow e \gamma$ process, giving
\begin{eqnarray}
\begin{array}{ll}
  \sqrt{|{\rm Re}(\delta^d_{12})_{LL/RR}^2|} 
    \simlt (10^{-2}\mbox{--}10^{-1}),\quad & 
  \sqrt{|{\rm Re}(\delta^d_{12})_{LL}(\delta^d_{12})_{RR}|} \simlt 10^{-3},
\\ \\
  \sqrt{|{\rm Im}(\delta^d_{12})_{LL/RR}^2|} 
    \simlt (10^{-3}\mbox{--}10^{-2}),\quad & 
  \sqrt{|{\rm Im}(\delta^d_{12})_{LL}(\delta^d_{12})_{RR}|} \simlt 10^{-4},
\end{array}
\label{eq:delta-exp_KK}
\end{eqnarray}
\begin{eqnarray}
\begin{array}{ll}
  |(\delta^u_{12})_{LL/RR}| \simlt (10^{-2}\mbox{--}10^{-1}),\quad & 
  |(\delta^u_{12})_{LL}| = |(\delta^u_{12})_{RR}| 
    \simlt (10^{-3}\mbox{--}10^{-2}),
\\ \\
  |(\delta^d_{13})_{LL/RR}| \simlt (0.1\mbox{--}1),\quad & 
  |(\delta^d_{13})_{LL}| = |(\delta^d_{13})_{RR}| \simlt 10^{-2},
\end{array}
\label{eq:delta-exp_DD-BB}
\end{eqnarray}
\begin{eqnarray}
  |(\delta^e_{12})_{LL}| \simlt (10^{-4}\mbox{--}10^{-3}),
\label{eq:delta-exp_LFV}
\end{eqnarray}
where we have kept only the bounds relevant to our model.  In deriving 
the above bounds, we have taken the gluino and squark masses to be 
the same order of magnitude with $m_{\tilde{q}} \simeq 500~{\rm GeV}$, 
and the same for the weak gaugino and slepton masses with $m_{\tilde{l}} 
\simeq 200~{\rm GeV}$.  For heavier superparticles, the bounds become 
weaker linearly with increasing superparticle masses, except for that on 
$|(\delta^e_{12})_{LL}|$, which scales quadratically with $m_{\tilde{l}}$.

In order to compare our model with the above bounds, we need to obtain 
the structure of the squark and slepton mass matrices at low energies. 
We first consider the flavor universal contribution.  It comes from 
two different sources.  The first is gauge mediation, generated at 
the scale $M_{\rm mess}$, while the other is a $U(1)_Y$ Fayet-Iliopoulos 
$D$-term piece, ${\rm Tr}(Y_{\tilde{f}} m_{\tilde{f}}^2) \neq 0$, of 
the soft masses generated at $M_*$, Eqs.~(\ref{eq:mu-mH2},~\ref{eq:mq2},%
~\ref{eq:ml2}).  The sfermion masses at a low energy, $\mu_R$, can 
then be written as
\begin{eqnarray}
  m_{\tilde{f}}^2(\mu_R) 
  &\simeq& 2 N_{\rm mess} 
    \sum_{a=1}^{3} C_a^{\tilde{f}} \frac{g_a^4(M_{\rm mess})}{(16\pi^2)^2} 
    \left[ 1 + \frac{N_{\rm mess}}{b_a} 
      \left( 1 - \frac{g_a^4(\mu_R)}{g_a^4(M_{\rm mess})} \right) \right] 
    \frac{F_X^2}{\langle X \rangle^2}
\nonumber\\
  && - \frac{6 Y_{\tilde{f}}}{5} \frac{g_1^2(\mu_R)}{16\pi^2} 
    \left( x_{H_u} - x_{H_d} + \frac{x_{Q_3} \alpha_q^2}{\tilde{y}} 
      -2 \frac{x_{U_3}}{\tilde{y}\, \alpha_q^2} 
      + \frac{x_{E_3} \alpha_\beta^2}{\tilde{y}\, \alpha_l^2} \right) 
    \frac{F_X^2}{\Lambda^2} \ln\frac{M_*}{\mu_R},
\label{eq:mf2-approx}
\end{eqnarray}
where $(b_1,b_2,b_3) = (33/5,1,-3)$ are the 321 beta-function coefficients, 
$Y_{\tilde{f}}$ represents hypercharges in the normalization that $Q$ 
has $Y_{\tilde{f}} = 1/6$, and $x_{H_u,H_d,Q_3,U_3,E_3}$ are the $O(1)$ 
factors in front of the corresponding soft masses generated at $M_*$. 
(Here, we have kept only the leading terms in $\epsilon$.)  As we 
will see in section~\ref{subsec:NLSP}, the $U(1)_Y$ $D$-term piece 
can considerably affect the superparticle spectrum, leading to 
interesting phenomenology.

The flavor violating elements of the sfermion mass matrices are 
renormalized among themselves, and are also generated from the 
flavor universal piece through the Yukawa couplings.  These 
effects, however, do not significantly modify the values of these 
elements in most of the parameter space.  We therefore take the 
approximation that the flavor non-universal part of the sfermion 
masses is parameterized by Eqs.~(\ref{eq:mq2},~\ref{eq:ml2}) with 
Eqs.~(\ref{eq:epsilon-QUD},~\ref{eq:epsilon-LE}) at low energies.%
\footnote{A possible contribution to $m_{\tilde{l}}^2$ from loops 
 of the right-handed neutrinos is also not important as long as 
 $(y_\nu)_{ij} \simlt O(1)$, which is the case for the $\epsilon$ 
 factor assignment of Eq.~(\ref{eq:epsilon-LE}) with $\alpha_l 
 \approx O(1)$.}
The chirality-preserving mass insertion parameters are then obtained 
by dividing these flavor violating elements by the (average) 
diagonal elements in the super-CKM basis.

With the low-energy mass parameters described above, one can study 
the constraints from flavor and $CP$ violation.  The scalar trilinear 
interactions in our model are generated only by renormalization group 
evolution, so that they are proportional to the corresponding Yukawa 
couplings with real proportionality constants, in the basis where 
the gaugino masses are real.  They, therefore, do not contribute 
to flavor or $CP$ violating processes.  The constraints on the 
$\alpha$ parameters are then obtained from Eqs.~(\ref{eq:delta-exp_KK}%
~--~\ref{eq:delta-exp_LFV}).  We find that for $\tilde{y} = 4\pi$ 
and $\epsilon = 0.05$, all constraints from the quark sector are 
satisfied, while $\mu \rightarrow e \gamma$ gives
\begin{equation}
  \alpha_l \simlt 1.8,
\label{eq:const-alpha-1}
\end{equation}
with no further constraints on $\alpha_q$ or $\alpha_\beta$. 
Taking $\epsilon = 0.1$, the constraints become stronger with 
both $\mu \rightarrow e \gamma$ and $K$-$\bar{K}$ mixing, giving
\begin{equation}
  \alpha_l \simlt 0.9,
\qquad
  \alpha_\beta \simlt 1.4.
\label{eq:const-alpha-2}
\end{equation}
These bounds are obtained for the superparticle mass scale of 
$m_{\tilde{l}} \sim 400~{\rm GeV}$, with $F_X/M_* \sim 1~{\rm TeV}$. 
(This corresponds to $m_{\tilde{q}} \sim 1.2~{\rm TeV}$, which is 
sufficient to avoid the LEP~II bound on the physical Higgs boson 
mass.)  While these bounds are rough ones, they show that there 
exists a consistent parameter region.  For heavier superparticles, 
the bounds become weaker and the region expands.

\subsection{The NLSP}
\label{subsec:NLSP}

As we have seen in section~\ref{subsec:SUSY-br}, the LSP is the 
gravitino with mass $\approx (0.1$~--~$10)~{\rm GeV}$.  In order to 
study phenomenology, it is important to determine which particles can 
be the NLSP.  Since the dominant contribution to the masses of most 
supersymmetric particles comes from gauge mediation, we first consider 
the spectrum without the corrections from tree-level pieces generated 
at $M_*$.  Since the masses are determined by the gauge charge, the 
lightest particles will be those neutral under $SU(3)_C$ and $SU(2)_L$. 
Therefore, the lightest gaugino is a neutralino, $\chi_1^0$ which is 
mostly bino, and the lightest sfermions are the right-handed sleptons. 
The mass of the bino at low energy is given by
\begin{equation}
  m_{\tilde{B}}(\mu_R) \simeq N_{\rm mess} 
    \frac{g_1^2(\mu_R)}{16\pi^2} \frac{F_X}{\langle X \rangle},
\label{eq:m_bino}
\end{equation}
while the mass of the sleptons can be derived from Eq.~(\ref{eq:mf2-approx}). 
From these two equations we see that with increasing $N_{\rm mess}$ the 
sleptons become lighter than the bino, while increasing $M_{\rm mess}$ 
makes the sleptons heavier because of renormalization group effects. 
Calculations show that for $N_{\rm mess} = 1$ the bino is always the 
NLSP, while for larger $N_{\rm mess}$ the sleptons can be lighter. 
In the case of $N_{\rm mess} = 3$ ($5$), for example, the sleptons are 
lighter than the bino for $M_{\rm mess} \simlt 10^{10}~(10^{12})~{\rm GeV}$. 

The bino mass in the present model is the same as in gauge 
mediation, but the slepton masses can deviate.  As discussed in 
section~\ref{subsec:flavor}, the sleptons receive the contribution 
from the $U(1)_Y$ $D$-term, indicated by the second line of 
Eq.~(\ref{eq:mf2-approx}).  This contribution is flavor universal 
so it does not affect the splitting among sleptons, but it 
affects the relation between the sleptons and the bino.  The other 
correction to gauge mediation comes from the tree-level masses in 
Eqs.~(\ref{eq:mq2},~\ref{eq:ml2}).  From Eqs.~(\ref{eq:epsilon-QUD},%
~\ref{eq:epsilon-LE}), we see that these mass terms are $\epsilon$ 
suppressed for most fields, but the effect can be $O(1)$ for 
$\tilde{\tau}_R$, and the unknown coefficient could even be negative 
as long as the sum of the tree-level and gauge mediated pieces bring 
the physical mass above direct detection bounds.  This means that 
$\tilde{\tau}_R$ could lie anywhere in the spectrum of $\tilde{e}_R$, 
$\tilde{\mu}_R$ and $\tilde{B}$.

The splitting between $\tilde{e}_R$ and $\tilde{\mu}_R$ is controlled 
almost entirely by the splitting at $M_*$ because the renormalization 
group running is universal up to small effects from the muon Yukawa 
coupling.  Phenomenology is governed by the splitting between mass 
eigenstates which is given by
\begin{equation}
  m_{\tilde{\mu}_R} - m_{\tilde{e}_R} 
  \approx \frac{m_{\tilde{\mu}_R}^2 - m_{\tilde{e}_R}^2} 
    {2\sqrt{m_{\tilde{e}_R,\tilde{\mu}_R}^2}} 
  \approx \frac{O(0.01)}{N_{\rm mess}^2} 
    \left(\frac{\alpha_\beta}{\alpha_l}\right)^2 
    \left(\frac{\Lambda/M_{\rm Pl}}{0.01}\right)^2 
    \frac{m_{\tilde B}^2}{\sqrt{m_{\tilde{e}_R,\tilde{\mu}_R}^2}}.
\label{eq:e-mu-split}
\end{equation}
The splitting between light generation sfermions is much larger than 
in the usual gauge mediation scenario.  It can be large enough that 
the heavier one can decay to the lighter by emission of an electron 
and a muon.

There are corners of parameter space where the NLSP is not a right-handed 
slepton or bino.  Since the contribution from the $U(1)_Y$ $D$-term in 
Eq.~(\ref{eq:mf2-approx}) has opposite signs for the left-handed and 
right-handed sleptons, it could invert the usual order between these 
two species.  The lighter stop could also be the NLSP because, like 
$\tilde{\tau}_R$, it has an $O(1)$ tree-level contribution to its mass. 
The stops also have a contribution from the large top Yukawa coupling, 
which decreases the masses through renormalization group evolution. 
While the tree-level piece is expected to be smaller than the $SU(3)_C$ 
gauge mediation piece, negative tree-level and Yukawa effects could 
combine to give a strongly interacting NLSP.  We do not consider these 
exotic NLSPs in the rest of this paper because they require large 
cancellation between independent effects.

\subsection{Proton decay}
\label{subsec:p-decay}

Dimension four proton decay in the present model can be forbidden by 
the $U(1)_H$ symmetry or matter parity.  Dimension five proton decay 
caused by colored Higgsino exchange is also absent because of the form 
of the Higgsino mass matrix determined by higher dimensional spacetime 
symmetry~\cite{Hall:2001pg}.  Proton decay in the present model can 
thus arise only from dimension six operators and cutoff suppressed 
dimension five operators.

As discussed in section~\ref{subsec:U1H}, we can take the charge 
assignment of $U(1)_H$ such that the operators $W \sim Q_i Q_j Q_k L_l$ 
and $U_i U_j D_k E_l$ are forbidden: $3q_Q+q_L \neq 0, -4$.  In this 
case, dimension five proton decay arises only from operators on the 
$y=\pi R$ brane which involve the X VEV.  The relevant interactions 
are $W \sim X^m Q_i Q_j Q_k L_l$ and $X^m U_i U_j D_k E_l$, which can 
be written for $3q_Q+q_L = -2m$ and $3q_Q+q_L = 2m-4$ ($m \in Z > 0$), 
respectively.  In the low-energy 4D effective theory, these interactions 
lead to dimension five operators
\begin{equation}
  W \approx 4\pi \epsilon_{Q_i} \epsilon_{Q_j} \epsilon_{Q_k} \epsilon_{L_l} 
      \frac{\Lambda^{m-1}}{M_{\rm Pl}^m} Q_i Q_j Q_k L_l
  \quad \mbox{and} \quad
    4\pi \epsilon_{U_i} \epsilon_{U_j} \epsilon_{D_k} \epsilon_{E_l} 
      \frac{\Lambda^{m-1}}{M_{\rm Pl}^m} U_i U_j D_k E_l,
\label{eq:eff-dim5-op}
\end{equation}
where the coefficients are evaluated using naive dimensional analysis, 
and we have used $\langle X \rangle \approx \Lambda^2/M_{\rm Pl}$. 
We find that the approximate sizes of these operators are obtained by 
replacing the colored Higgsino mass by $4\pi M_{\rm Pl}^m/\Lambda^{m-1}$ 
in the corresponding expressions in the minimal supersymmetric $SU(5)$ 
grand unified theory.  The resulting proton decay rate is thus much 
smaller than the current experimental bound for all the values of 
$3q_Q+q_L \neq 0, -4$.

Dimension six operators are generated in the present model only through 
brane localized terms, since without them exchange of bulk gauge bosons 
does not transform a quark into a lepton or vice versa.  (Note that 
different 321 multiplets arise from different $SU(5)$ multiplets, see 
Eqs.~(\ref{eq:bc-T}~--~\ref{eq:bc-O}).)  The relevant terms are kinetic 
mixing operators $K \sim T^\dagger T'$, $T^\dagger T''$, $F^\dagger F'$ 
and cutoff suppressed dimension six operators $K \sim T^\dagger T' 
T^\dagger T''$, $T^\dagger T' F^{\prime\dagger} F$ on the $y=0$ brane. 
Here, we have omitted factors involving the gauge multiplet needed 
to make operators gauge invariant, and the existence of Hermitian 
conjugates is implied.  The kinetic mixing terms lead, through unified 
gauge boson exchange, to dimension six operators at low energies, 
whose coefficients have approximately the size obtained by replacing 
the unified gauge boson mass by $1/\pi R$ in the corresponding 
minimal supersymmetric $SU(5)$ expressions.  For $1/\pi R \approx 
10^{15}~{\rm GeV}$, this leads to a proton decay rate somewhat larger 
than the current experimental bound~\cite{Shiozawa:1998si}.  This 
implies that the compactification scale should be somewhat larger (by 
a factor of a few) or the coefficients of the original kinetic mixing 
operators should be suppressed (by an order of magnitude or so).  This 
potential difficulty does not arise in weakly coupled models, an example 
of which will be discussed in section~\ref{sec:weakly-coupled}.  The 
coefficients of low-energy dimension six operators arising from 
the cutoff suppressed operators are similar in size to those in 
the minimal supersymmetric $SU(5)$ model, so that they do not 
lead to proton decay at a dangerous level.

In summary, proton decay in the present model is caused by dimension 
six operators, originating from terms on the $y=0$ brane.  Since the 
wavefunction values for the first and second generation fields on this 
brane are typically of the same order, the proton can decay into final 
states containing $\mu^+$ with a similar rate to those containing 
$e^+$.  This provides interesting signatures for future proton 
decay experiments.

\subsection{Precision gauge coupling unification}
\label{subsec:gcu}

Strongly coupled grand unification in higher dimensions allows a precise 
calculation for gauge coupling unification~\cite{Nomura:2001tn,Hall:2001xb}. 
Incalculable corrections arising from the cutoff scale physics are 
suppressed, and the corrections from higher dimensional fields between 
the energy interval between $M_*$ and $1/\pi R$ are precisely calculated. 
Here we study this issue in the model of section~\ref{sec:model}.

We phrase the degree of the success of gauge coupling unification in 
terms of the prediction of $\alpha_s(M_Z) = g_3^2(M_Z)/4\pi$ obtained 
from $g_{1,2}(M_Z)$, where $g_{1,2,3}$ represent the standard model 
gauge couplings.  In particular, we consider the deviation of the 
prediction in the present model, $\alpha_s^{\rm 5D}$, from that obtained 
by assuming the exact unification in the MSSM, $\alpha_s^{\rm SGUT,0}$:
\begin{equation}
  \delta\alpha_s \equiv \alpha_s^{\rm 5D} - \alpha_s^{\rm SGUT,0} 
    \simeq -\frac{1}{2\pi} \alpha_s^2 \Delta.
\label{eq:delta-alpha}
\end{equation}
Here, $\Delta$ parameterizes corrections from higher dimensional 
fields, which can be calculated within higher dimensional effective 
field theory.  Using the result of Ref.~\cite{Choi:2003ff}, we find 
that in the present model
\begin{eqnarray}
  \Delta &=& -\frac{3}{7} \ln(\pi R M_*) 
    -3 \ln(\epsilon_{Q_1} \epsilon_{Q_2} \epsilon_{Q_3})
    +\frac{15}{7} \ln(\epsilon_{U_1} \epsilon_{U_2} \epsilon_{U_3})
\nonumber\\
  && +\frac{9}{7} \ln(\epsilon_{D_1} \epsilon_{D_2} \epsilon_{D_3})
  -\frac{9}{7} \ln(\epsilon_{L_1} \epsilon_{L_2} \epsilon_{L_3})
  +\frac{6}{7} \ln(\epsilon_{E_1} \epsilon_{E_2} \epsilon_{E_3}),
\label{eq:delta}
\end{eqnarray}
where we have used the approximation that the Higgs doublets are strictly 
localized to the $y=\pi R$ brane.  (The term $-(9/7)\ln(\epsilon_{H_u} 
\epsilon_{H_d})$ should be added to the right-hand-side if the Higgs 
fields are delocalized.)  Inserting Eqs.~(\ref{eq:epsilon-QUD},%
~\ref{eq:epsilon-LE}) into this equation, we obtain
\begin{equation}
  \Delta = -\frac{3}{7} \ln(\pi R M_*) -\frac{135}{7} \ln\alpha_q 
    - \frac{45}{7}\ln\alpha_l + \frac{45}{7}\ln\alpha_\beta.
\label{eq:delta-2}
\end{equation}
Considering that the logarithms are expected to be of order unity, we 
find that $\Delta$ is typically of $O(10)$, with the sign depending on 
the values of $\alpha_{q,l,\beta}$.  For typical superparticle spectra, 
including the one considered here, a good fit to the experimental 
values of $g_{1,2,3}(M_Z)$ is obtained for
\begin{equation}
  \Delta^{\rm exp} \approx 5 \pm O(1).
\label{eq:Delta-exp}
\end{equation}
The expression in our model, Eq.~(\ref{eq:delta-2}), can easily 
accommodate this value.

\subsection{Collider signatures}
\label{subsec:signature}

Phenomenology of the general flavorful supersymmetry scenario has 
been discussed in Ref.~\cite{Nomura:2007ap}.  Here we summarize some 
of the basic features in the context of the present model.  As we saw 
in section~\ref{subsec:NLSP}, this model has a large portion of parameter 
space where there is a charged NLSP which is stable for the purposes 
of collider studies.  Unlike the conventional scenarios, the NLSP 
in flavorful supersymmetry could be a $\tilde{\tau}_R$ or a right-handed 
slepton of a different flavor.  Heavy stable charged particles are 
relatively easy to see at colliders.  By measuring their velocity and 
momentum, their mass can be deduced.  The mass of the charged NLSP be 
can measured to better than $1\%$ at the LHC by measuing only a few 
hundred NLSPs with $0.6 < \beta < 0.91$~\cite{Ambrosanio:2000ik}.

Once the NLSP mass is known, it is possible to fully reconstruct events 
even in the hadronic environment of the LHC.  Therefore we can determine 
the flavor content of the NLSP by taking its invariant mass with other 
leptons in the event.  If the NLSP is found to be mostly selectron or 
smuon, this is definitive evidence for nontrivial flavor structure in 
the supersymmetry breaking sector, and possibly for flavorful supersymmetry. 
In addition, once we learn the dominant flavor of the NLSP, we can look 
for NLSP production in association with leptons of other flavors to 
measure the mixing angles of the NLSP.

Because the lifetime of the NLSP is quite long, it can be studied 
in a cleaner environment. One proposal involves using the muon 
tracker to determine where in the surrounding rock an NLSP went, 
and extracting pieces of rock that likely contain NLSPs to study 
them elsewhere~\cite{De Roeck:2005bw}.  Another possibility is 
to build a large stopper detector outside of one of the main 
detectors which can stop the NLSPs and then measure the decay 
products~\cite{Hamaguchi:2006vu}.  This would allow precise 
measurements of the lifetime of the NLSP as well as the masses 
of the decay products.  As pointed out in Ref.~\cite{Nomura:2007ap}, 
a particularly distinct signature of flavorful supersymmetry is 
monochromatic electrons or muons in the decay of the NLSP, indicating 
a two body decay of a selectron or smuon.  This is not a possibility 
in the conventional scenarios because the $\tilde{\tau}_R$ is the NLSP, 
and it decays to a $\tau$ which further decays, so the many body decay 
causes the leptons to have a broad spectrum.  Even if the NLSP is 
a $\tilde{\tau}_R$, a stopper detector will allow us to look for rare 
decays into other flavors and precisely measure the flavor content 
of the NLSP.  The stopper detector can also check to see if the 
LSP is the gravitino.  From the kinematics, the mass of the LSP 
can be measured, which can then be tested against the supergravity 
prediction which relates the lifetime of the NLSP to the mass of 
the gravitino~\cite{Buchmuller:2004rq}.

While the signatures are much more spectacular if there is a slepton 
NLSP, evidence for flavorful supersymmetry can still be found with 
a neutralino NLSP.  One possibility is to look for direct slepton 
production from Drell-Yan processes and measure the spectrum through 
kinematic variables such as $M_{T2}$~\cite{Lester:1999tx}.  This 
is difficult because it requires high statistics and the Drell-Yan 
cross section falls rapidly with increasing slepton mass.  Another 
possibility is to look for multiple edges in flavor-tagged dilepton 
invariant mass distributions as in Ref.~\cite{Bartl:2005yy}.  This 
will allow us to find different flavors of sleptons if they are 
separated by more than a few GeV, which we would expect in flavorful 
supersymmetry.  Finally, we could also study the spectrum of left-handed 
sleptons or even squarks to look for flavor non-universality.  While 
these measurements are more difficult than those with stable sleptons, 
they could still provide information on the flavor structure of the 
supersymmetry breaking sector.

\section{4D Realization --- Model in Warped Space}
\label{sec:warped}

The model in section~\ref{sec:model} has been formulated in flat space, 
but we can also consider a similar model in warped space, along the 
lines of Ref.~\cite{Nomura:2006pn}.  An interesting feature of this 
model is that it allows for a 4D interpretation through the AdS/CFT 
correspondence, providing a picture of realizing flavorful supersymmetry 
in a 4D setup.

Specifically, we take the metric
\begin{equation}
  ds^2 = e^{-2ky} \eta_{\mu\nu} dx^\mu dx^\nu + dy^2,
\label{eq:metric}
\end{equation}
where $k$ denotes the inverse curvature radius of the warped spacetime. 
The two branes are located at $y=0$ (the UV brane) and $y=\pi R$ (the 
IR brane).  The scales of these branes are chosen to be $k \approx 
10^{17}~{\rm GeV}$ and $k' \equiv k\, e^{-\pi kR} \approx 10^{16}~{\rm 
GeV}$, respectively.  The cutoff scale of the 5D theory is taken to 
be $M_* \approx 10^{18}~{\rm GeV}$.  The gauge symmetry structure is 
as described in section~\ref{sec:model}; the bulk $SU(5)$ symmetry 
is broken to 321 on the IR brane at $y=\pi R$.  The IR brane thus 
serves the role of breaking the unified symmetry.

The configuration of the matter and Higgs fields is as described 
in section~\ref{sec:model}.  The locations of these fields are 
controlled by the bulk masses, and the resulting Yukawa couplings 
are given by Eq.~(\ref{eq:Yukawa}), where the $\epsilon$ factors 
are given by Eq.~(\ref{eq:epsilon}) with $M_\Phi \rightarrow M_\Phi 
- k/2$.  The analysis of $U(1)_H$ and supersymmetry breaking is 
as in sections~\ref{subsec:PQ}~--~\ref{subsec:U1H-breaking}.  (Note 
that the cutoff scale on the IR brane is warped down to $M'_* \equiv 
M_* e^{-\pi kR} \approx 10^{17}~{\rm GeV}$.)  This leads to phenomenology 
discussed in sections~\ref{subsec:flavor}, \ref{subsec:NLSP} and 
\ref{subsec:signature}.  Dimension four and five proton decay is 
negligible for the reasons described in section~\ref{subsec:p-decay}. 
Dimension six proton decay is also not dangerous as the unified 
gauge boson mass is now of order $\pi k' \approx 10^{16}~{\rm GeV}$. 
For gauge coupling unification, we can show, using the results 
of~\cite{Choi:2002ps}, that the threshold correction is still given 
by the formula Eq.~(\ref{eq:delta}).  (Note that the contribution 
from the Higgs doublets to differential running shuts off above 
$M'_*$, since these fields are localized on the IR brane.)  The 
experimental values of the low-energy gauge couplings are thus 
successfully reproduced, as seen in section~\ref{subsec:gcu}.

The model described here has the following 4D interpretation through 
the AdS/CFT correspondence.  At very high energies above $k' \approx 
10^{16}~{\rm GeV}$, the theory is a 4D supersymmetric $SU(5) \times G$ 
gauge theory, where $SU(5)$ is the unified gauge group and $G$ some 
quasi-conformal gauge group.  There are three generations of matter 
fields, $3 \times ({\bf 10} + {\bf 5}^*)$ of $SU(5)$ (and possibly 
three right-handed neutrinos), but not the Higgs fields.  There are 
also fields charged under $G$, some of which are charged under $SU(5)$ 
as well.  At the scale $k' \approx 10^{16}~{\rm GeV}$, the $G$ sector 
deviates from the conformal fixed point, breaking the unified $SU(5)$ 
symmetry to 321 by the gauge dynamics.  It also produces the MSSM 
Higgs doublets and the supersymmetry breaking sector containing 
$X$ as composite states.  The effective theory below $k'$ is thus 
the MSSM (and possibly three right-handed neutrinos) together with 
the supersymmetry breaking sector.

An important point is that the interaction strengths of the matter 
fields to the $G$ sector are controlled by the dimensions of operators 
coupling matter to fields charged under $G$.  In general, these 
dimensions are generation dependent.  Moreover, since $G$ is strongly 
interacting above $k'$, the anomalous dimensions for these operators 
can be large.  As a result, the interaction strengths of matter to 
the $G$ sector strongly vary between different generations, and since 
the Higgs doublets and $X$ arise as composite states of $G$, the 
interactions of matter to these states show strong generation dependence. 
Since the origin of this generation dependence is common for the matter 
couplings to the Higgs fields (the Yukawa couplings) and to the $X$ 
field (supersymmetry breaking couplings), the patterns of these two 
classes of couplings are correlated.  The correlation is exactly 
the one given in Eqs.~(\ref{eq:Yukawa},~\ref{eq:int-M-X}), realizing 
flavorful supersymmetry.

We have considered here a 4D theory in which the $G$ sector is 
quasi-conformal and has a large 't~Hooft coupling above the dynamical 
scale, motivated by the warped space construction.  The dynamics 
described above, however, are independent of these assumptions. 
The same dynamics can also be incorporated, in principle, in a purely 
4D theory whose 't~Hooft coupling is not necessarily large above $k'$. 
The quasi-conformal nature of the dynamics is also not essential. 
It will be interesting to construct an explicit example of purely 
4D theory in which the $G$ sector exhibits different renormalization 
group behavior, e.g. asymptotic freedom, above the dynamical scale 
$\Lambda_G \approx 10^{16}~{\rm GeV}$.

\section{Weakly Coupled (Non-Unified) Models}
\label{sec:weakly-coupled}

In this section we present a non-unified model of flavorful supersymmetry 
in higher dimensions.  Here we do not require that the theory is strongly 
coupled at the cutoff scale, nor that it possesses the $U(1)_H$ symmetry. 
Rather, we assume that certain operators are small at the cutoff scale 
due to ultraviolet physics.

We consider a supersymmetric $SU(3)_C \times SU(2)_L \times U(1)_Y$ 
gauge theory in 5D flat spacetime, compactified on an $S^1/Z_2$ orbifold: 
$0 \leq y \leq \pi R$.  As in the model of section~\ref{sec:model}, 
the two Higgs doublets are localized towards the $y=\pi R$ brane, where 
supersymmetry is broken by the $F$-term VEV of a chiral superfield 
$X$.  The matter fields are introduced in the bulk as hypermultiplets, 
whose zero modes $Q_i, U_i, D_i, L_i, E_i$ (and $N_i$) are identified 
with the MSSM matter fields.  The wavefunction profiles of the zero 
modes are controlled by the bulk masses $M_\Phi$ ($\Phi = Q_i, U_i, 
D_i, L_i, E_i, N_i$), as seen in section~\ref{subsec:fermion-masses}.

\begin{figure}
\begin{center}
  \input{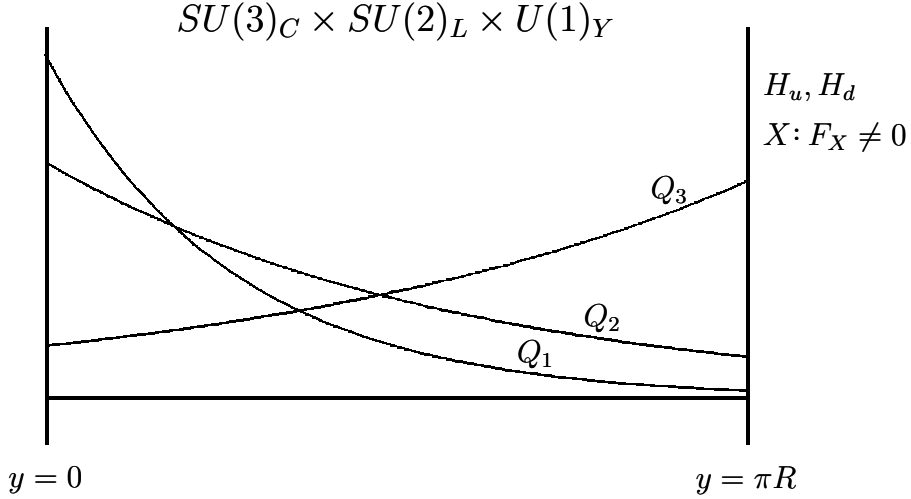}
\caption{A schematic depiction of the configuration for various fields.}
\label{fig:config-2}
\end{center}
\end{figure}

We do not require that the theory is strongly coupled at the cutoff 
scale $M_*$, which is taken to be a factor of a few above $1/R$.  We 
then naturally expect that the operators located on branes have $O(1)$ 
coefficients in units of $M_*$.  This leads to the 4D Yukawa couplings 
of Eq.~(\ref{eq:4D-Yukawa}) with
\begin{equation}
  (y_u)_{ij} \approx \epsilon_{Q_i} \epsilon_{U_j},
\qquad
  (y_d)_{ij} \approx \epsilon_{Q_i} \epsilon_{D_j},
\qquad
  (y_e)_{ij} \approx \epsilon_{L_i} \epsilon_{E_j},
\qquad
  (y_\nu)_{ij} \approx \epsilon_{L_i} \epsilon_{N_j},
\label{eq:Yukawa-2}
\end{equation}
at low energies, where $\epsilon_\Phi$ are given by Eq.~(\ref{eq:epsilon}). 
By choosing $\epsilon_\Phi$ and $\tan\beta$ to be as given in 
Eqs.~(\ref{eq:epsilon-QUD}~--~\ref{eq:tan-beta-gen}) with $\tilde{y} = 1$, 
this reproduces the gross structure of the observed quark and lepton 
masses and mixings, Eqs.~(\ref{eq:q-l-masses},~\ref{eq:q-l-mixings}).%
\footnote{Here we have assumed that the Majorana masses for $N_i$ are 
 on the $y = \pi R$ brane, but not on the $y = 0$ brane.  This can be 
 realized, for example, by introducing the $U(1)_{B-L}$ symmetry broken 
 on the $y = \pi R$ brane.}
The configuration of the matter fields, as well as those of the Higgs 
and supersymmetry breaking fields, are depicted schematically in 
Fig.~\ref{fig:config-2}.

The supersymmetry breaking parameters are generated through the 
interactions of the MSSM states to the $X$ field on the $y=\pi R$ 
brane.  In the absence of the $U(1)_H$ symmetry, the superpotential 
operators $W \sim X Q_i U_j H_u + X Q_i D_j H_d + X L_i E_j H_d 
+ X L_i N_j H_u$ are not forbidden in general.  These operators generate 
flavor non-universal left-right mixing terms for the squarks and 
sleptons that require relatively heavy superparticles to avoid the 
constraints from low-energy flavor and $CP$ violating processes. 
Here we assume that these operators are somehow suppressed.  We also 
assume that the direct $\mu$ term, $W \sim H_u H_d$, is absent. 
Note that these assumptions are technically natural because of the 
nonrenormalization theorem.  The supersymmetry breaking parameters 
are then generated by the K\"ahler potential operators and ${\cal L} 
\sim \int\!d^2\theta\, X {\cal W}_a^\alpha {\cal W}_{a\alpha} + 
{\rm h.c.}$, where ${\cal W}_a^\alpha$ ($a=1,2,3$) are the 321 
gauge field strength superfields, giving
\begin{equation}
  M_a \approx \mu \approx \frac{F_X}{M_*}, 
\qquad
  m_{H_u}^2 \approx m_{H_d}^2 \approx B\mu 
    \approx \left(\frac{F_X}{M_*}\right)^2,
\label{eq:Ma-Higgs}
\end{equation}
\begin{equation}
  (m_{\tilde{q}}^2)_{ij} \approx 
    \epsilon_{Q_i} \epsilon_{Q_j} \left(\frac{F_X}{M_*}\right)^2,
\qquad
  (m_{\tilde{u}}^2)_{ij} \approx 
    \epsilon_{U_i} \epsilon_{U_j} \left(\frac{F_X}{M_*}\right)^2,
\qquad
  (m_{\tilde{d}}^2)_{ij} \approx 
    \epsilon_{D_i} \epsilon_{D_j} \left(\frac{F_X}{M_*}\right)^2,
\label{eq:mq2-2}
\end{equation}
\begin{equation}
  (m_{\tilde{l}}^2)_{ij} \approx 
    \epsilon_{L_i} \epsilon_{L_j} \left(\frac{F_X}{M_*}\right)^2,
\qquad
  (m_{\tilde{e}}^2)_{ij} \approx 
    \epsilon_{E_i} \epsilon_{E_j} \left(\frac{F_X}{M_*}\right)^2,
\label{eq:ml2-2}
\end{equation}
\begin{equation}
  (a_u)_{ij} 
    \approx \bigl\{ (y_u)_{kj} (\eta_Q)_{ki} + (y_u)_{ik} (\eta_U)_{kj} 
      + (y_u)_{ij} \bigr\}\, \frac{F_X}{M_*},
\label{eq:a_u}
\end{equation}
\begin{equation}
  (a_d)_{ij} 
    \approx \bigl\{ (y_d)_{kj} (\eta_Q)_{ki} + (y_d)_{ik} (\eta_D)_{kj} 
      + (y_d)_{ij} \bigr\}\, \frac{F_X}{M_*},
\label{eq:a_d}
\end{equation}
\begin{equation}
  (a_e)_{ij} 
    \approx \bigl\{ (y_e)_{kj} (\eta_L)_{ki} + (y_e)_{ik} (\eta_E)_{kj} 
      + (y_e)_{ij} \bigr\}\, \frac{F_X}{M_*}.
\label{eq:a_e}
\end{equation}
Here, we have omitted $O(1)$ coefficients in each term, and 
$(\eta_\Phi)_{ij} \approx \epsilon_{\Phi_i} \epsilon_{\Phi_j}$ 
($\Phi = Q,U,D,L,E$) are general complex $3 \times 3$ matrices. 
This gives a correlation between the Yukawa couplings 
Eq.~(\ref{eq:Yukawa-2}), and the supersymmetry breaking 
parameters Eqs.~(\ref{eq:Ma-Higgs}~--~\ref{eq:a_e}), realizing 
flavorful supersymmetry.  Note that because of the absence of 
a factor $4\pi$ in Eq.~(\ref{eq:Yukawa-2}), the mass splittings 
between different generation sfermions in Eqs.~(\ref{eq:mq2-2},%
~\ref{eq:ml2-2}) can be larger than those in the strongly 
coupled case.

The model has other flavor violating contributions to the supersymmetry 
breaking parameters, but they can be controlled.  For example, 
loops of the higher dimensional gauge fields produce flavor violating 
supersymmetry breaking masses at $1/R$, but they are not much larger 
than the tree-level masses in the parameter region considered, as long 
as the coefficients of the matter brane kinetic operators at $y=0$ 
are of order $1/16\pi^2 M_*$ or smaller.  Note that this size of the 
coefficients is technically natural.  The matter 4-point K\"ahler 
potential operators on the $y=0$ brane also give flavor violating 
contributions at loop level.  They are, however, suppressed by 
a factor of $1/(\pi R M_*)^5$ and negligible.  Possible contributions 
from bulk higher dimension operators are also expected to be small.

The compactification scale $1/R$ in the present model is naturally of 
order the unification scale $M_U \approx 10^{16}~{\rm GeV}$ to preserve 
the successful supersymmetric prediction for the low-energy gauge couplings. 
In this case, the gaugino and sfermion masses are of order $\tilde{m} 
\approx F_X/M_U$ while the gravitino mass is $m_{3/2} \approx 
F_X/M_{\rm Pl}$, so that $m_{3/2} \approx (M_U/M_{\rm Pl}) \tilde{m} 
\approx (1~\mbox{--}~10)~{\rm GeV}$, leading to signatures discussed 
in section~\ref{subsec:signature} with the NLSP being one of the 
right-handed sleptons.  The compactification scale, however, can in 
principle take any value larger than of order a few TeV, in which 
case the gravitino may be (much) lighter.  Note that the supersymmetry 
breaking parameters of Eqs.~(\ref{eq:Ma-Higgs}~--~\ref{eq:a_e}) are 
running parameters evaluated at the scale $1/R$.  The low-energy 
superparticle spectrum is obtained by evolving them down to the 
weak scale using renormalization group equations.

Here we have presented a non-unified model of flavorful supersymmetry 
in 5D.  It is, however, straightforward to make it a unified model, 
e.g., based on $SU(5)$.  We simply have to adopt the field content 
and boundary conditions of section~\ref{subsec:SU5-5D} and follow 
the analysis above.  To understand gauge coupling unification, we 
need to assume that incalculable brane-localized gauge kinetic terms 
on the $y=\pi R$ brane are somehow suppressed (or universal), but 
dangerous proton decay can be easily suppressed, possibly by $U(1)_R$ 
symmetry~\cite{Hall:2001pg}.%
\footnote{It is interesting to note that the 321 gaugino masses do 
 not have to be unified at the unification scale even if the model 
 is unified because the gaugino mass operators reside on the $y=\pi R$ 
 brane, where the active gauge group is only 321~\cite{Hall:2001pg}.}
It is also straightforward to extend the model to higher dimensions. 
The only requirement is that the Higgs fields and the supersymmetry 
breaking field $X$ are localized in the same place in the extra dimensions.%
\footnote{To be more precise, it is sufficient to require that the matter 
interactions to the Higgs and $X$ fields are suppressed by common 
wavefunction factors, allowing the Higgs and $X$ to propagate in 
different subspaces.}
An advantage of such a setup is that we can suppress cutoff scale 
dimension-five proton decay operators by localizing the $Q,U,E$ and 
$D,L$ fields in different subspaces in higher dimensions.  These 
extensions allow us to realize flavorful supersymmetry in a wide 
variety of higher dimensional models, with varying spacetime 
dimensions, compact space geometries, and gauge groups.

\section{Conclusions}
\label{sec:concl}

In this paper we have presented explicit models of flavorful 
supersymmetry in higher dimensions.  The basic idea is to localize 
the Higgs fields and the supersymmetry breaking field in the same 
location in the extra dimension(s).  The interactions of matter 
fields to the Higgs fields (the Yukawa couplings) and to the 
supersymmetry breaking field (operators generating the supersymmetry 
breaking parameters) then receive the same suppression factors 
from the wavefunction profiles of the matter fields.  This leads 
to a specific correlation between these two classes of interactions, 
realizing flavorful supersymmetry.  The resulting phenomenology 
at future colliders is very rich, while stringent experimental 
constraints from the low-energy flavor and $CP$ violating 
processes can all be satisfied.

We have constructed a unified model of flavorful supersymmetry in 
5D, in which the theory is strongly coupled at the cutoff scale. 
Supersymmetry breaking is mediated to the supersymmetric standard 
model sector by a combination of cutoff suppressed operators and 
gauge mediation.  This model addresses various issues in supersymmetric 
unification.  We have also presented a model in warped space, which 
allows us to obtain a picture of realizing flavorful supersymmetry 
in a 4D setup, through the AdS/CFT correspondence.  Finally, we 
have discussed models which do not require that the theory is 
strongly coupled at the cutoff scale.  This construction can be 
easily extended to a wide variety of higher dimensional theories, 
with varying spacetime geometries and gauge groups.

It is interesting to note that the present setup is very generic 
in the context of a single extra dimension.  If we want to explain 
the observed hierarchical structure of the Yukawa couplings by 
wavefunction overlaps between the matter and Higgs fields, the 
simplest way is to localize the Higgs fields to one of the branes 
and lighter generation matter more towards the other brane.  Now, 
if the supersymmetry breaking field $X$ is {\it not} localized to 
the same brane as the Higgs fields, interactions of lighter generation 
matter to $X$ are {\it not} suppressed, leading to large flavor 
violating supersymmetry breaking masses.  To avoid this problem, 
we need to localize $X$ to the same brane as the Higgs fields 
(unless some other flavor universal mediation mechanism dominates). 
This gives the spectrum of flavorful supersymmetry.

As the LHC will turn on this year, it is important to explore possible 
theoretical constructions and experimental signatures of supersymmetric 
theories.  The models presented here provide an example in which 
the supersymmetry breaking spectrum can be a window into the physics 
of flavor in the standard model.  If supersymmetry is discovered at 
the LHC, it will be interesting to see if the longstanding assumption 
of flavor universality holds, or if there is a richer flavor structure 
within the supersymmetry breaking sector.  This structure could give 
us information about the physics of flavor which could lie at energy 
scales as high as the unification or Planck scale.

\section*{Acknowledgment}

This work was supported in part by the U.S. DOE under Contract 
DE-AC02-05CH11231, and in part by the NSF under grant PHY-04-57315. 
The work of Y.N. was also supported by the NSF under grant PHY-0555661, 
by a DOE OJI, and by the Alfred P. Sloan Research Foundation. 
The work of D.S. was supported by the Alcatel-Lucent Foundation.


\begin{thebibliography}{99}

\bibitem{Feng:2007ke}
J.~L.~Feng, C.~G.~Lester, Y.~Nir and Y.~Shadmi,
arXiv:0712.0674 [hep-ph].

\bibitem{Nomura:2007ap}
Y.~Nomura, M.~Papucci and D.~Stolarski,
arXiv:0712.2074 [hep-ph].

\bibitem{Kitano:2006ws}
R.~Kitano and Y.~Nomura,
arXiv:hep-ph/0606134.

\bibitem{ArkaniHamed:1999dc}
N.~Arkani-Hamed and M.~Schmaltz,
Phys.\ Rev.\  D {\bf 61}, 033005 (2000)
[arXiv:hep-ph/9903417];
T.~Gherghetta and A.~Pomarol,
Nucl.\ Phys.\  B {\bf 586}, 141 (2000)
[arXiv:hep-ph/0003129].

\bibitem{Kaplan:2000av}
D.~E.~Kaplan and T.~M.~P.~Tait,
JHEP {\bf 0006}, 020 (2000)
[arXiv:hep-ph/0004200];
JHEP {\bf 0111}, 051 (2001)
[arXiv:hep-ph/0110126].

\bibitem{Hall:2002ci}
L.~J.~Hall and Y.~Nomura,
Phys.\ Rev.\  D {\bf 66}, 075004 (2002)
[arXiv:hep-ph/0205067].

\bibitem{Abe:2004tq}
H.~Abe, K.~Choi, K.~S.~Jeong and K.~i.~Okumura,
JHEP {\bf 0409}, 015 (2004)
[arXiv:hep-ph/0407005].

\bibitem{Kawamura:2000ev}
Y.~Kawamura,
Prog.\ Theor.\ Phys.\  {\bf 105}, 999 (2001)
[arXiv:hep-ph/0012125].

\bibitem{Hall:2001pg}
L.~J.~Hall and Y.~Nomura,
Phys.\ Rev.\  D {\bf 64}, 055003 (2001)
[arXiv:hep-ph/0103125].

\bibitem{Hall:2001zb}
L.~J.~Hall, Y.~Nomura and D.~R.~Smith,
Nucl.\ Phys.\  B {\bf 639}, 307 (2002)
[arXiv:hep-ph/0107331];
L.~J.~Hall, J.~March-Russell, T.~Okui and D.~R.~Smith,
JHEP {\bf 0409}, 026 (2004)
[arXiv:hep-ph/0108161].

\bibitem{Hebecker:2002re}
A.~Hebecker and J.~March-Russell,
Phys.\ Lett.\  B {\bf 541}, 338 (2002)
[arXiv:hep-ph/0205143].

\bibitem{Nomura:2001tn}
Y.~Nomura,
Phys.\ Rev.\  D {\bf 65}, 085036 (2002)
[arXiv:hep-ph/0108170].

\bibitem{Hall:2001xb}
L.~J.~Hall and Y.~Nomura,
Phys.\ Rev.\  D {\bf 65}, 125012 (2002)
[arXiv:hep-ph/0111068].

\bibitem{Ibe:2007km}
M.~Ibe and R.~Kitano,
arXiv:0705.3686 [hep-ph].

\bibitem{Nomura:2007cc}
Y.~Nomura and M.~Papucci,
arXiv:0709.4060 [hep-ph].

\bibitem{Giudice:1988yz}
G.~F.~Giudice and A.~Masiero,
Phys.\ Lett.\  B {\bf 206}, 480 (1988);
U.~Ellwanger,
Phys.\ Lett.\  B {\bf 133} (1983) 187.

\bibitem{Dine:1981gu}
M.~Dine and W.~Fischler,
Phys.\ Lett.\ B {\bf 110}, 227 (1982);
Nucl.\ Phys.\ B {\bf 204}, 346 (1982);
L.~Alvarez-Gaum\'{e}, M.~Claudson and M.~B.~Wise,
Nucl.\ Phys.\ B {\bf 207}, 96 (1982);
S.~Dimopoulos and S.~Raby,
Nucl.\ Phys.\ B {\bf 219}, 479 (1983).

\bibitem{Dine:1994vc}
M.~Dine, A.~E.~Nelson and Y.~Shirman,
Phys.\ Rev.\ D {\bf 51}, 1362 (1995)
[arXiv:hep-ph/9408384];
M.~Dine, A.~E.~Nelson, Y.~Nir and Y.~Shirman,
Phys.\ Rev.\ D {\bf 53}, 2658 (1996)
[arXiv:hep-ph/9507378].

\bibitem{ArkaniHamed:2001tb}
See, e.g.,
N.~Arkani-Hamed, T.~Gregoire and J.~G.~Wacker,
JHEP {\bf 0203}, 055 (2002)
[arXiv:hep-th/0101233].

\bibitem{Manohar:1983md}
A.~Manohar and H.~Georgi,
Nucl.\ Phys.\  B {\bf 234}, 189 (1984);
Z.~Chacko, M.~A.~Luty and E.~Ponton,
JHEP {\bf 0007}, 036 (2000)
[arXiv:hep-ph/9909248].

\bibitem{Hall:1999sn}
See, e.g.,
L.~J.~Hall, H.~Murayama and N.~Weiner,
Phys.\ Rev.\ Lett.\  {\bf 84}, 2572 (2000)
[arXiv:hep-ph/9911341];
T.~Yanagida and J.~Sato,
Nucl.\ Phys.\ Proc.\ Suppl.\  {\bf 77}, 293 (1999)
[arXiv:hep-ph/9809307];
P.~Ramond,
Nucl.\ Phys.\ Proc.\ Suppl.\  {\bf 77}, 3 (1999)
[arXiv:hep-ph/9809401].

\bibitem{Kitano:2006wz}
R.~Kitano,
Phys.\ Lett.\  B {\bf 641}, 203 (2006)
[arXiv:hep-ph/0607090].

\bibitem{Giudice:1997ni}
G.~F.~Giudice and R.~Rattazzi,
Nucl.\ Phys.\  B {\bf 511}, 25 (1998)
[arXiv:hep-ph/9706540].

\bibitem{Ibe:2007gf}
M.~Ibe and R.~Kitano,
arXiv:0710.3796 [hep-ph].

\bibitem{Peccei:1977hh}
R.~D.~Peccei and H.~R.~Quinn,
Phys.\ Rev.\ Lett.\  {\bf 38}, 1440 (1977).

\bibitem{Hall:1985dx}
L.~J.~Hall, V.~A.~Kostelecky and S.~Raby,
Nucl.\ Phys.\  B {\bf 267}, 415 (1986).

\bibitem{Gabbiani:1996hi}
F.~Gabbiani, E.~Gabrielli, A.~Masiero and L.~Silvestrini,
Nucl.\ Phys.\  B {\bf 477}, 321 (1996)
[arXiv:hep-ph/9604387];
A.~Masiero, S.~K.~Vempati and O.~Vives,
arXiv:0711.2903 [hep-ph],
and references therein.

\bibitem{Shiozawa:1998si}
M.~Shiozawa {\it et al.}  [Super-Kamiokande Collaboration],
Phys.\ Rev.\ Lett.\  {\bf 81}, 3319 (1998)
[arXiv:hep-ex/9806014].

\bibitem{Choi:2003ff}
K.~w.~Choi, I.~W.~Kim and W.~Y.~Song,
Nucl.\ Phys.\  B {\bf 687}, 101 (2004)
[arXiv:hep-ph/0307365].

\bibitem{Ambrosanio:2000ik}
S.~Ambrosanio, B.~Mele, S.~Petrarca, G.~Polesello and A.~Rimoldi,
JHEP {\bf 0101}, 014 (2001)
[arXiv:hep-ph/0010081].

\bibitem{De Roeck:2005bw}
A.~De Roeck, J.~R.~Ellis, F.~Gianotti, F.~Moortgat, K.~A.~Olive and L.~Pape,
Eur.\ Phys.\ J.\  C {\bf 49}, 1041 (2007)
[arXiv:hep-ph/0508198].

\bibitem{Hamaguchi:2006vu}
K.~Hamaguchi, M.~M.~Nojiri and A.~de Roeck,
JHEP {\bf 0703}, 046 (2007)
[arXiv:hep-ph/0612060];
K.~Hamaguchi, Y.~Kuno, T.~Nakaya and M.~M.~Nojiri,
Phys.\ Rev.\  D {\bf 70}, 115007 (2004)
[arXiv:hep-ph/0409248];
J.~L.~Feng and B.~T.~Smith,
Phys.\ Rev.\  D {\bf 71}, 015004 (2005)
[Erratum-ibid.\  D {\bf 71}, 0109904 (2005)]
[arXiv:hep-ph/0409278].

\bibitem{Buchmuller:2004rq}
W.~Buchm\"{u}ller, K.~Hamaguchi, M.~Ratz and T.~Yanagida,
Phys.\ Lett.\  B {\bf 588}, 90 (2004)
[arXiv:hep-ph/0402179].

\bibitem{Lester:1999tx}
C.~G.~Lester and D.~J.~Summers,
Phys.\ Lett.\  B {\bf 463}, 99 (1999)
[arXiv:hep-ph/9906349].

\bibitem{Bartl:2005yy}
A.~Bartl, K.~Hidaka, K.~Hohenwarter-Sodek, T.~Kernreiter, W.~Majerotto and W.~Porod,
Eur.\ Phys.\ J.\  C {\bf 46}, 783 (2006)
[arXiv:hep-ph/0510074];
G.~L.~Bayatian {\it et al.}  [CMS Collaboration],
J.\ Phys.\ G {\bf 34} (2007) 995.

\bibitem{Nomura:2006pn}
Y.~Nomura, D.~Poland and B.~Tweedie,
JHEP {\bf 0612}, 002 (2006)
[arXiv:hep-ph/0605014].

\bibitem{Choi:2002ps}
K.~w.~Choi and I.~W.~Kim,
Phys.\ Rev.\  D {\bf 67}, 045005 (2003)
[arXiv:hep-th/0208071];
W.~D.~Goldberger, Y.~Nomura and D.~R.~Smith,
Phys.\ Rev.\  D {\bf 67}, 075021 (2003)
[arXiv:hep-ph/0209158].

\end{thebibliography}
\end{document}